\documentclass[onecolumn]{pasj02} 
\usepackage[switch,mathlines]{lineno} 
\usepackage{natbib} 
\input{pasjsup.sty}
\usepackage[dvipsnames]{xcolor}

\jyear{2025}
\Received{}
\Accepted{}

\begin{document} 

\title{Disk instability model incorporating a variable inner disk radius in SS Cyg and U Gem}

\author{
 Mariko \textsc{Kimura}\altaffilmark{1}\altemailmark\orcid{0009-0002-1729-8416} \email{mariko.kimura@se.kanazawa-u.ac.jp} 
 and 
 Yoji \textsc{Osaki}\altaffilmark{2}
}
\altaffiltext{1}{Advanced Research Center for Space Science and Technology, College of Science and Engineering, Kanazawa University, Kakuma, Kanazawa, Ishikawa 920-1192}
\altaffiltext{2}{Department of Astronomy, School of Science, University of Tokyo, Hongo, Tokyo 113-0033}




\maketitle

\begin{abstract}

Previous theoretical studies indicate that the inner disk in dwarf novae evaporates into a high-temperature, optically thin, and geometrically thick accretion flow during quiescence, with the inner edge moving toward the white dwarf at the onset of an outburst. 
We incorporate this process into the numerical model developed by \citet{kim23sscyg} and test the code on two representative dwarf novae, SS~Cyg and U~Gem. 
By modeling the inner accretion flow, we calculate the optical, ultraviolet (UV), and X-ray luminosities. 
Our results show that evaporation suppresses the inside-out outbursts without requiring a radially dependent viscosity parameter in the cold state. 
The observed time delay between the rise in UV luminosity and the onset of the optical outburst is more than one day, which is successfully reproduced when the inner disk is truncated at several $\times 10^{9}$~cm in the standard evaporation model.
However, while the modeled accretion rate at the inner disk edge in U~Gem accounts for its quiescent X-ray luminosity, the rate in SS~Cyg remains insufficient. 
This discrepancy in SS~Cyg suggests that SS~Cyg may require either more efficient evaporation or an additional mass supply into the coronal cavity via gas-stream overflow. 
By accounting for disk evaporation, our simulations offer a refined version of the disk instability model for dwarf nova outbursts that naturally explains the observed multiwavelength light curves.

\end{abstract}


\section{Introduction}

Dwarf novae (DNe) are a subclass of cataclysmic variables (CVs) characterized by intermittent outbursts with amplitudes of 2--5~mag in the optical band and recurrence times ranging from a few weeks to several months. 
CVs are close binary systems with orbital periods of typically a few hours, in which a Roche-lobe-filling low-mass secondary star transfers mass to an accreting white dwarf (WD). 
An accretion disk is formed around the WD (see \citealt{war95book} for a comprehensive review of CVs).

The intermittent outbursts of DNe are now understood to arise from sudden brightenings of the accretion disk, triggered by the thermal-viscous instability operating within the disk (see \citealt{osa96review}). 
This mechanism, commonly referred to as the thermal limit-cycle instability, is described by the S-shaped thermal equilibrium curve relating the disk surface density to effective viscosity. 
The curve contains an unstable branch around ${\sim}$10$^{4}$~K, caused by partial hydrogen ionization, and bracketed by two stable branches: a hot branch with fully ionized hydrogen and high accretion rates, and a cold branch with neutral hydrogen and low accretion rates \citep{hos79DImodel,mey81DNoutburst}. 
The disk alternates between these two stable states, and the resultant heating and cooling fronts propagate across the disk. 
While the mass-transfer rate from the secondary star remains nearly constant, the mass-accretion rate through the disk drastically varies as a consequence of this instability. 
Numerous one-dimensional numerical simulations based on this framework have successfully reproduced DN outbursts (see a recent review by \citealt{ham20review}). 

We have recently resumed such numerical simulations with the aim of applying them to a variety of DN phenomena. 
As a first attempt, we investigated the effects of tilted accretion disks in IW Andromedae-type DNe using the numerical code developed by \citet{kim20tiltdiskmodel}. 
Although that work focused on tilted disks, the same code can be applied to non-tilted systems. 
Indeed, we examined the anomalous outbursts of SS Cyg using the non-tilted version of the code with some improvements in \citet{kim23sscyg}.

Although these codes allow for a variable outer disk radius, they assume a fixed inner disk radius at the WD surface. 
Implementing a variable inner disk edge is important because several observational issues appear to require truncation of the inner disk. One such issue is the ``UV delay'', in which the rise in UV flux lags behind the rise in optical flux at outburst onset \citep{whe03sscyg,lon96ugem}.
In addition, the inner disk is likely truncated during quiescence; otherwise, the predicted quiescent X-ray luminosity would be far lower than the observed value if the disk extended down to the WD surface (see e.g., the blue line in the top panel of Fig.~4 in \citealt{kim23sscyg}).
A further long-standing issue is the suppression of frequent, low-amplitude inside-out outbursts in simulations. 
If the disk always extends to the WD surface, such outbursts commonly appear, despite not being observed \citep{min86DNDI,ham98diskmodel}. 
\textcolor{black}{This issue is particularly noteworthy in WZ Sge-type stars because their outburst interval spans decades \citep{kat15wzsge}}.
To suppress inside-out outbursts in the disk extending to the WD surface, a radial dependence of the viscosity parameter in the cold branch was necessarily introduced (see, \citealt{min89quiescenceviscosity}).
Disk truncation during quiescence is a promising solution to these problems \textcolor{black}{as suggested by \citet{ham97wzsgemodel} and \citet{min98wzsge}}.

When the accretion rate is sufficiently low in quiescence, the inner disk may fail to remain an optically thick cold disk and instead evaporate into a high-temperature, optically thin, and geometrically thick coronal flow \citep{mey94siphonflow}. 
This coronal flow, with characteristic temperatures of ${\sim}$10$^{8}$~K, emits hard X-rays \citep{pat85CVXrayemission1}. 
Incorporating such evaporation into numerical simulations should therefore produce optical, UV, and X-ray light curves more consistent with observations.

In this paper, we implement a time-dependent disk inner radius in our numerical code based on \citet{kim23sscyg}, an improved version of the original code by \citet{kim20tiltdiskmodel}. 
Truncation of the inner disk and time-dependent disk inner radius have been treated by \cite{men00BHXN,dub01XNmodel}, but their works aimed at X-ray transient sources. 
While \citet{ham99UVdelay} recognized evaporation as a key mechanism for the UV delay, their analysis did not include this effect in numerical simulations.
The truncation of the inner disk by the magnetic field of the WD was treated by \citet{ham17IPoutburst} and \citet{ham20modeling} in their simulations. In this paper, we examine the truncation of the inner disk by evaporation for the non-magnetic CVs.
To achieve this, we implement a simplified formulation of evaporation and refilling of the inner disk to our numerical code, model the optical, UV, and X-ray emission, and compute their light curves.
Our goal is to test whether introducing a variable inner disk edge can resolve several observational issues, including the UV delay, the quiescent X-ray luminosity, and the frequency of inside-out outbursts. 
We apply our simulations to SS Cyg and U Gem, two well-known prototype DNe above the period gap.

The structure of this paper is as follows. 
Section~2 describes our numerical method for implementing a variable disk inner edge and the choice of model parameters. 
The simulation results and comparisons with observational data are presented in Section~3. 
We discuss the implications of our findings in Section~4 and summarize our conclusions in Section~5.

\section{Method of numerical simulations}\label{sec:3}

\subsection{Implementation of variable disk inner edge}\label{ssec:31}

\begin{figure}[htbp]
 \begin{center}
  \includegraphics[width=8cm]{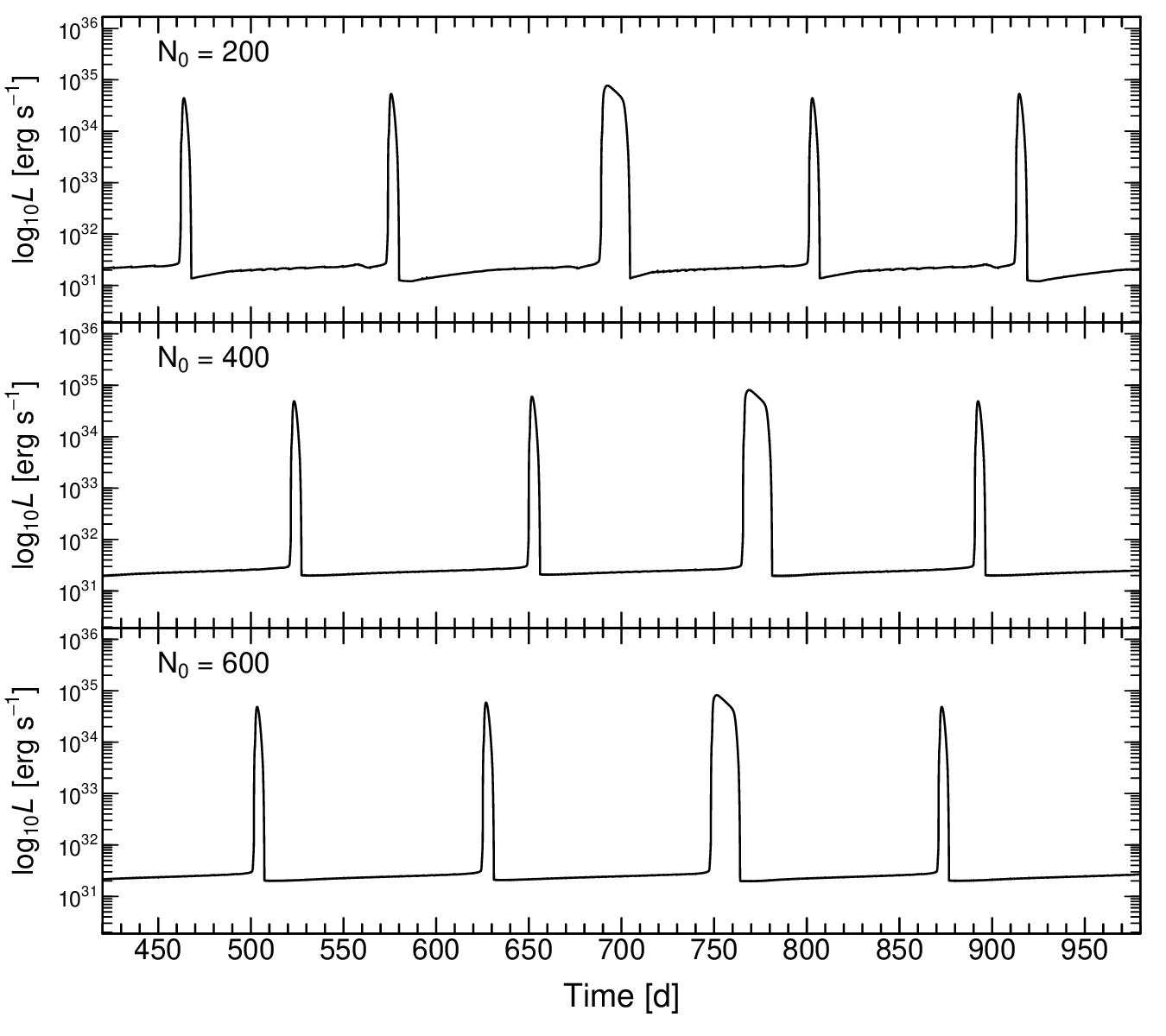} 
 \end{center}
\caption{Comparison of simulated light curves for models with $N_0$ = 200, 400, and 600 \textcolor{black}{for U Gem}. Each run incorporates a variable disk inner edge with the standard evaporation rate of $\dot{M}_{\rm base}$ = 10$^{15}$~g~s$^{-1}$. 
{Alt text: Three rows of panels. X axis shows the time \textcolor{black}{from 420 to 980 days}. Y axis shows \textcolor{black}{the bolometric disk luminosity} on a logarithmic scale from 10 to 30.5 to 10 to 36 ergs per second. The top, middle, and bottom panels show the simulation light curve with 200, 400, and 600 meshes, respectively. These light curves include 7 outbursts. The number of short outbursts between two long outbursts is 2.}
}
\label{fig:N0-comparison}
\end{figure}

The numerical simulation method employed here is essentially the same as that in \citet{kim23sscyg} except for the treatment of the disk inner edge.
The numerical code used in \citet{kim23sscyg} is a revised version of that in \citet{kim20tiltdiskmodel}, which is in turn based on \citet{ich92diskradius}.
We formulate mass, angular momentum, and energy conservation laws for the accretion disk in the radial direction, and solve
them by a hybrid method of explicit and implicit integration.
As for the treatment of the outer disk edge, we assume that the expansion halts at the tidal truncation radius due to the strong tidal torques and the disk cannot expand further when the disk reaches the tidal truncation radius.
The radial mesh points are equally spaced in $\sqrt{r}$ and the maximum number of concentric annuli, denoted here by $N_0$, is 400.
We compared the resultant light curves of our numerical simulations with $N_0$ = 200, 400, and 600 and confirmed they did not depend on $N_0$ very much, though the case of $N_0$ = 200 cannot generate smooth light curves during quiescence \textcolor{black}{and though the outburst intervals are slightly shorter than those in the other cases} (see Figure \ref{fig:N0-comparison}).
\textcolor{black}{Here we note that $L$ in this figure represents the sum of the radiation flux from each annulus, which is called ``the bolometric disk luminosity'' in this paper.}

\begin{figure}[htbp]
 \begin{center}
  \includegraphics[width=8cm]{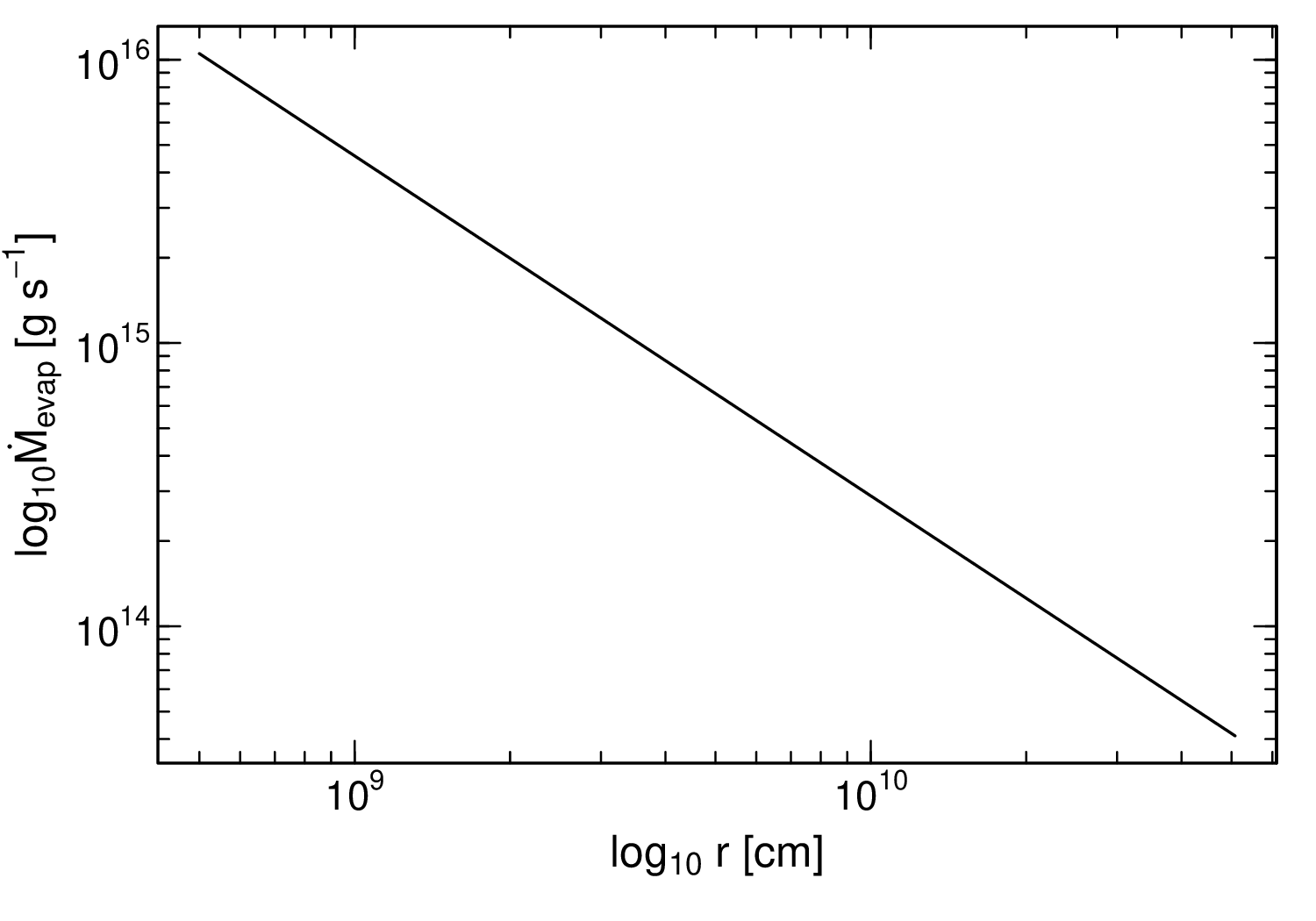} 
 \end{center}
\caption{Evaporation rate $\dot{M}_{\rm evap}$ as a function of radius for parameters of SS Cyg with $M_1 = 0.94 M_{\odot}$, $\lambda = 0.25$, and $\dot{M}_{\rm base}$=10$^{15}$~g~s$^{-1}$.
{Alt text: A single plot. X axis shows the distance from the white dwarf on a logarithmic scale from 5 times 10 to 8 to 5 times 10 to 10 centimeters. Y axis shows the evaporation rate on a logarithmic scale from 4 times 10 to 14 to 10 to 16 gram per second.}
}
\label{fig:mdot-evap}
\end{figure}

To model the variable inner edge of the accretion disk due to evaporation, we adopt an approach fundamentally similar to that of \citet{men00BHXN}. 
\textcolor{black}{Specifically, we assume a radial dependence for the evaporation rate from the disk, denoted as $\dot{M}_{\mathrm{evap}}(r)$. 
The functional form employed is based on the prescription proposed by \citet{mey99evap}.
According to \citet{mey94siphonflow} and \citet{mey99evap}, some of the evaporated mass is lost into a coronal wind, and the evaporation rate is the sum of the wind mass loss rate and the accretion rate to the WD via the coronal flow.
\citet{mey99evap} provided not the evaporation rate itself but the accretion rate.
Here we define $\lambda$ as the fraction of mass loss via coronal winds relative to the total mass supplied from the disk to the corona through evaporation, and then, $\dot{M}_{\mathrm{evap}}(r)$ is expressed as}
\begin{equation}
  \dot{M}_{\rm evap} =\textcolor{black}{ \left(\frac{\dot{M}_{\rm base}}{1 - \lambda}\right)} \frac{( M_1 / M_{\odot} )^{2.4}}{(r / 10^{9.5})^{1.2}},
\label{mdot-evap}
\end{equation}
where $M_1$ represents the WD mass.
The parameter $\dot{M}_{\rm base}$ characterizes the strength of evaporation. 
Following \citet{mey94siphonflow} and \citet{mey99evap}, we adopt $\dot{M}_{\rm base} = 10^{15}$~g~s$^{-1}$ and $\lambda = 0.25$ for their standard values. 
The adopted radial profile of the evaporation rate used in our simulations is illustrated in Figure \ref{fig:mdot-evap}.
The evaporation rate decreases monotonically with radius. 

We further assume that evaporation occurs within a narrow radial zone near the inner disk edge, such that all mass loss from this region is attributed to evaporation and it occurs from the inner edge of the disk. 
Accordingly, we impose the condition:
\begin{equation}
\dot{M}_{\rm in}(R_{\rm in}) = \dot{M}_{\rm evap}(R_{\rm in}), 
\label{evaporation}
\end{equation} 
where $R_{\rm in}(t)$ denotes the time-dependent radius of the inner disk edge.

\textcolor{black}{
To incorporate this evaporation model into our simulation code, we proceed as follows. 
The code operates in two stages: (1) advancing the dynamical variables (mass and angular momentum) by one time step (from ``old'' to ``new''), and (2) solving the thermal equations implicitly for the updated variables. 
Evaporation affects only the first stage.}

\textcolor{black}{
To handle the time-dependent inner edge during the first stage, we divide the process into an intermediate step and a final update. 
Initially, we tentatively advance the system to $t_{\rm new} = t_{\rm old} + dt$ under the assumption that the inner edge remains fixed at its position from $t_{\rm old}$. 
We denote the intermediate quantities computed in this step with a superscript $\#$. 
In this intermediate step, we calculate the mass and the angular momentum in each mesh and the mass accretion rate $\dot{M}$ at mesh interfaces.  
For the innermost mesh (the innermost annulus), this yields the tentative mass loss rate at the inner boundary $\dot{M}^{\#}_{\rm in}$ and the tentative angular momentum of the innermost mesh $\Delta J^{\#}_{\rm new}$. 
If the inner edge coincides with the WD surface, $\dot{M}^{\#}_{\rm in}$ corresponds to the accretion rate onto the WD.
We also store the mass and the angular momentum of each mesh for the purpose of their later use.
}

\textcolor{black}{
When evaporation is active and the inner edge varies with time, we must modify the variables. 
We apply this modification only to the innermost mesh; variables and interfaces in all other meshes remain unchanged. 
The innermost mesh is bounded by $r_{k-1}$ and $r_k$, where $k \ge 1$ is the mesh index. 
The mesh center is $r_{k-1/2} = \frac{r_{k-1} + r_k}{2}$, and the inner edge lies at $R_{\rm in} = r_{k-1}$.
}

\textcolor{black}{
First, the tentative mass of the innermost mesh at the new time step, $\Delta M^{\#}_{\mathrm{new}}(r_{k-1/2})$, computed under the fixed inner boundary assumption, is given by:
\begin{equation}
\Delta M^{\#}_{\mathrm{new}}(r_{k-1/2}) = \Delta M(r_{k-1/2}, t_{\mathrm{old}}) + (\dot{M}(r_k) -\dot{M}^{\#}_{\mathrm{in}}(r_{k-1})) dt, 
\label{mass-in-new-fix}
\end{equation} 
where $\Delta M$ denotes the amount of mass in the mesh. 
We then correct this intermediate state to account for the actual evaporation. By replacing the tentative inner boundary mass loss ($\dot{M}^{\#}_{\mathrm{in}}$) with the evaporation rate  ($\dot{M}_{\mathrm{evap}}$), the final updated mass of the innermost mesh, 
 $\Delta M_{\mathrm{new}}$, becomes:
\begin{equation}
\begin{split}
\Delta M_{\mathrm{new}}(r_{k-1/2}) &= \Delta M(r_{k-1/2}, t_{\mathrm{old}}) + (\dot{M}(r_k) - \dot{M}_{\mathrm{evap}}(r_{k-1})) dt \\ 
&= \Delta M^{\#}_{\mathrm{new}}(r_{k-1/2}) - (\dot{M}_{\mathrm{evap}}(r_{k-1}) - \dot{M}^{\#}_{\mathrm{in}}(r_{k-1})) dt.
\end{split}
\label{mass-in-new}
\end{equation}
Here we use equation (\ref{evaporation}). 
}

\textcolor{black}{
Similarly, we update the angular momentum of the innermost mesh at the new time step, $\Delta J_{\mathrm{new}}(r_{k-1/2})$, 
by adjusting the intermediate value $\Delta J^{\#}_{\mathrm{new}}(r_{k-1/2})$ for the difference in mass loss:
\begin{equation}
\Delta J_{\mathrm{new}}(r_{k-1/2}) = \Delta J^{\#}_{\mathrm{new}}(r_{k-1/2}) - (\dot{M}_{\mathrm{evap}} (r_{k-1})- \dot{M}^{\#}_{\mathrm{in}}(r_{k-1}))~h_{\mathrm{old}}(r_{k-1})~dt, 
\label{ang-in-new}
\end{equation} 
where  $h(r_i)$ represents the specific angular momentum defined by $\sqrt{G M_1 r_i}$.
}

\textcolor{black}{
Since $\Delta J^{\#}_{\mathrm{new}}(r_{k-1/2}) = \Delta M^{\#}_{\mathrm{new}}(r_{k-1/2}) h_{\mathrm{old}}(r_{k-1/2})$, we can rewrite $\Delta J^{\#}_{\mathrm{new}}(r_{k-1/2})$ by using equation (\ref{mass-in-new}) as: 
\begin{equation}
\Delta J^{\#}_{\mathrm{new}}(r_{k-1/2}) = \Delta M_{\mathrm{new}}(r_{k-1/2})~h_{\mathrm{old}}(r_{k-1/2}) + (\dot{M}_{\mathrm{evap}} (r_{k-1})- \dot{M}^{\#}_{\mathrm{in}}(r_{k-1}))~h_{\mathrm{old}}(r_{k-1/2})~dt.
\label{Jnew-fix}
\end{equation}
By substituting equation (\ref{Jnew-fix}) into equation (\ref{ang-in-new}), we obtain:
\begin{equation}
\Delta J_{\mathrm{new}}(r_{k-1/2}) = \Delta M_{\mathrm{new}}(r_{k-1/2}) ~h_{\mathrm{old}}(r_{k-1/2}) + (\dot{M}_{\mathrm{evap}}(r_{k-1}) - \dot{M}^{\#}_{\mathrm{in}}(r_{k-1}))~(h_{\mathrm{old}}(r_{k-1/2}) - h_{\mathrm{old}}(r_{k-1}))~dt.
\label{ang-in-new-2}
\end{equation}
Since $\Delta J_{\mathrm{new}}(r_{k-1/2}) = \Delta M_{\mathrm{new}}(r_{k-1/2}) h_{\mathrm{new}}(r_{k-1/2})$, we obtain $h_{\mathrm{new}}(r_{k-1/2})$ by dividing both sides of equation (\ref{ang-in-new-2}) by $\Delta M_{\mathrm{new}}(r_{k-1/2})$:
\begin{equation}
h_{\mathrm{new}}(r_{k-1/2}) = h_{\mathrm{old}}(r_{k-1/2}) + \frac{(\dot{M}_{\mathrm{evap}}(r_{k-1}) - \dot{M}^{\#}_{\mathrm{in}}(r_{k-1}))~(h_{\mathrm{old}}(r_{k-1/2}) - h_{\mathrm{old}}(r_{k-1}))~dt}{\Delta M_{\mathrm{new}} (r_{k-1/2})}.
\label{hin-new}
\end{equation}
}

\textcolor{black}{
This is the very equation that determines the innermost radius at the new time step.
We can see from equation (\ref{hin-new}) that the radius of the inner edge increases (i.e., the coronal cavity expands) if $\dot{M}_{\mathrm{evap}} (r_{k-1}) > \dot{M}^{\#}_{\mathrm{in}}(r_{k-1})$
and it decreases (i.e., the coronal cavity shrinks) if $\dot{M}_{\mathrm{evap}}(r_{k-1}) < \dot{M}^{\#}_{\mathrm{in}}(r_{k-1})$. 
}

The total mass and angular-momentum conservation laws in equations (A7) and (A8) of \citet{kim20tiltdiskmodel} are also modified as follows:
\begin{eqnarray}
   M_{\rm disk}(t_{\rm new}) = M_{\rm disk}(t_{\rm old}) + (\dot{M}_{\rm tr} - \dot{M}_{\rm evap})~dt~\textcolor{black}{= \sum_{i=k}^{N} \Delta M_{i-1/2}}, \\
   J_{\rm disk}(t_{\rm new}) = J_{\rm disk}(t_{\rm old}) + (h_{\rm LS} \dot{M}_{\rm tr} - h_{\rm old}(r_{k-1})\dot{M}_{\rm evap} - D_{\rm total})~dt~\textcolor{black}{= \sum_{i=k}^{N} h_{i-1/2} \Delta M_{i-1/2}},
\label{conservations}
\end{eqnarray}
where $M_{\rm disk}$, $J_{\rm disk}$, $\dot{M}_{\rm tr}$, $h_{\rm LS}$, and \textcolor{black}{$D_{\rm total}$} represent the total mass of the disk, the total angular momentum of the disk, the mass transfer rate, the specific angular momentum at the Lubow-Shu radius, and \textcolor{black}{the total tidal torque exerted on the accretion disk (see equation (A10) of \citealt{kim20tiltdiskmodel})}, respectively.

\textcolor{black}{
We have confirmed that our numerical scheme preserves the exact conservation of mass and angular momentum in the disk throughout the time evolution by verifying the equality between the second and the third expressions in equations (9) and (\ref{conservations}).}

Once the first stage is finished, we proceed to the second stage in solving the thermal equations to obtain the temperature of each mesh.  
During an outburst, the heating front propagates inward and eventually reaches the inner edge. The inner edge moves inward and it finally reaches the WD surface.
After the decline phase of the outburst, $\dot{M}_{\rm evap}$ exceeds $\dot{M}_{\rm in}$, and then, the inner edge recedes from the WD surface.
During quiescence, the \textcolor{black}{coronal cavity} continues to expand, but its expansion rate gradually slows.
At some points, a next outburst is triggered, and $\dot{M}_{\rm in}$ exceeds $\dot{M}_{\rm evap}$.
Then the \textcolor{black}{coronal cavity} is refilled by the accreted gas and the inner edge finally reaches the WD surface.

We split the innermost mesh into two if its width exceeds 1.5 times the original width, assigning both new meshes the same temperature as before splitting. 
The mass is distributed assuming equal surface density in the two new meshes. 
We merge the innermost mesh with its neighbor mesh if its width becomes smaller than half of its original width; the temperature of the merged mesh is determined such a way to conserve the thermal energy.
These procedures follow the same method used for splitting and merging the outermost mesh.

\subsection{Calculation of optical luminosity}\label{ssec:32}

To compare our simulation results with observational characteristics, we calculate the optical, UV, and X-ray fluxes for each target.
The optical flux in each system originates from the accretion disk (including the bright spot), the WD, and the secondary star.
We calculated the optical $V$-band magnitude of the disk by the same method as that used in \citet{dub18DItest}.
The contribution from the bright spot is included in the same manner as described in \citet{kim20tiltdiskmodel}.
Additionally, we include the flux contributions from the WD and the secondary star.
The binary parameters of SS Cyg and U Gem are summarized in Table \ref{binary-parameters}.
We assume that the emission from the WD and the secondary star can be described by a single-temperature blackbody; we then apply the filtering method of \citet{dub18DItest}, to estimate their respective $V$-band magnitudes.

\subsection{Modeling of optically-thick boundary layer: extreme UV flux during outburst}\label{ssec:33}

\begin{figure*}
 \begin{center}
 \begin{minipage}{0.325\hsize}
  \includegraphics[width=5.5cm]{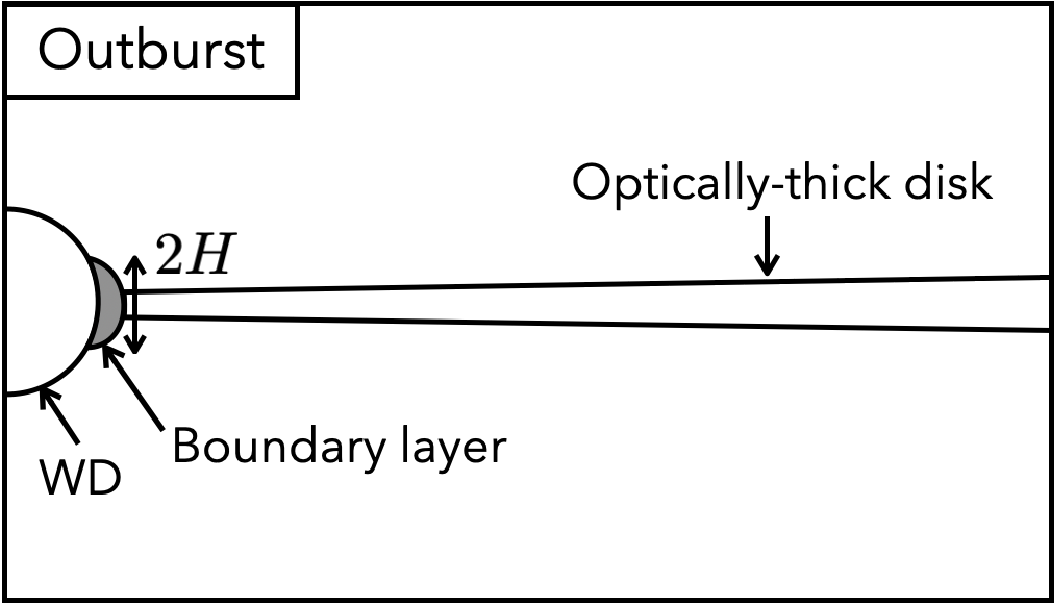} 
\end{minipage}
\begin{minipage}{0.325\hsize}
  \includegraphics[width=5.5cm]{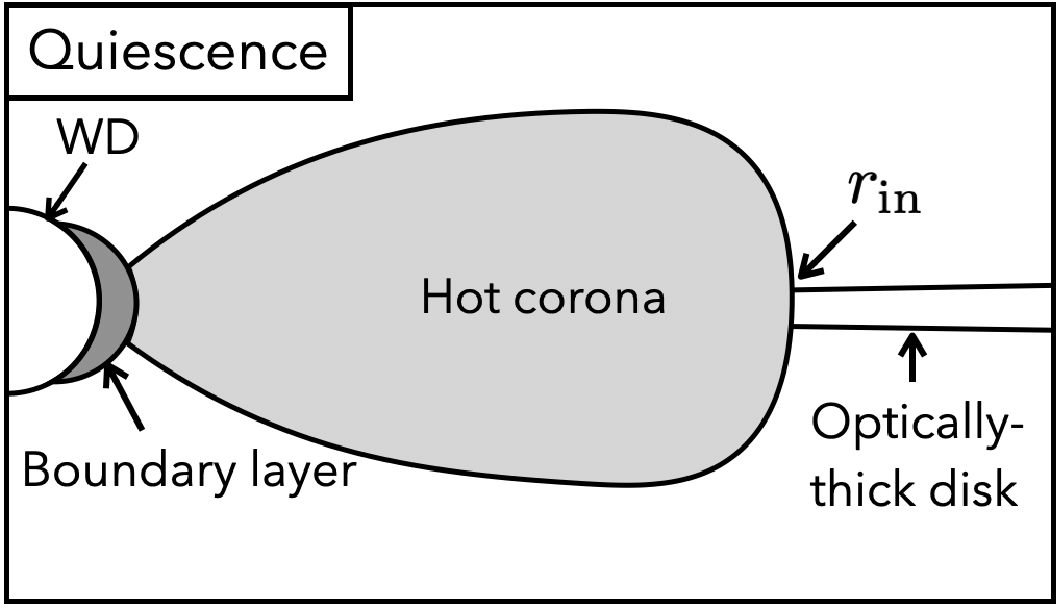}
\end{minipage}
\begin{minipage}{0.325\hsize}
  \includegraphics[width=5.5cm]{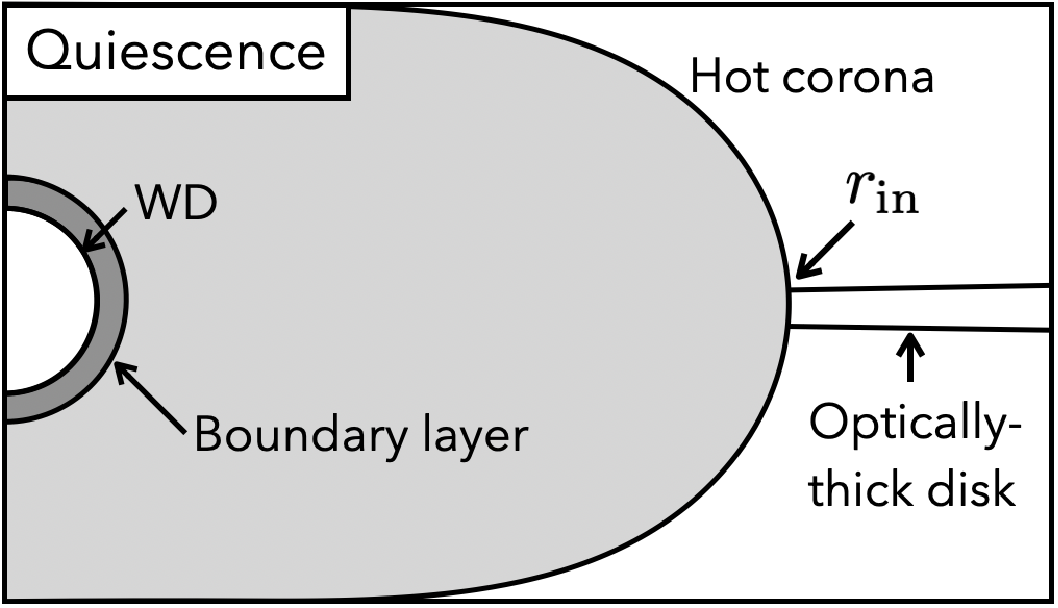}
\end{minipage}
\end{center}
\caption{Schematic edge-on views of the inner accretion flow geometries considered in this study. (Left) The disk during outburst, where an optically thick BL bridges the WD surface and the optically thick accretion disk. (Middle) The disk in quiescence, featuring an optically thin BL attached to the WD surface and the coronal flow extending between the BL and the truncated optically thick disk. (Right) An alternative quiescent configuretion, in which the coronal flow expands nearly spherically due to inefficient radiative cooling, resulting in an optically thin BL that envelops the entire WD surface.
{Alt text: Three columns of diagrams.}
}
\label{fig:inner-flow-picture}
\end{figure*}

In DNe, the majority of the extreme UV (EUV) flux is thought to originate from the boundary layer (BL) between the accretion disk and the WD during outbursts. 
During quiescence, the mass accretion rate to the BL is low; the gas in the BL is heated to approximately the virial temperature and emits hard X-ray. 

Once an outburst occurs, the mass accretion rate to the BL increases rapidly. 
At a certain point, the BL transitions from the optically thin, hot state to an optically thick state, subsequently emitting EUV radiation. 
The critical accretion rate for this transition is believed to be approximately $\dot{M}_{\rm acc}=10^{16}$~g~s$^{-1}$, although its exact value remains uncertain. 
In this paper we assume that the transition from an optically thin BL to an optically thick one occurs when the inner disk edge reaches the WD surface. 
Conversely, we assume that the opposite transition occurs when the inner disk edge recedes from the WD surface as an outburst ends. 

During quiescence, the UV sources are limited to the WD. 
In contrast, during an outburst, the mass accretion rate is high and the inner disk becomes UV-bright as it transforms into an optically thick and geometrically thin disk.
The BL, where the rotational energy of the accreted gas is released as radiation, also becomes optically thick and functions as an additional EUV source in this case (see the left panel of Figure \ref{fig:inner-flow-picture}). 
We compute the total EUV luminosity by summing the contributions from the inner disk, the WD, and the optically thick BL. 

We assume that the BL is steady.
The energy loss rate at the BL is given by
\begin{equation}
  \dot{E}_{\rm loss} = \frac{1}{2} \dot{M} {R_1}^2 {\Omega_{\rm K}}^2 \left(1 - \frac{{\Omega_{\rm WD}}^2}{{\Omega_{\rm K}}^2}\right) \approx \frac{1}{2} \dot{M} {R_1}^2 {\Omega_{\rm K}}^2,
\label{Edot-loss}
\end{equation}
where $\Omega_{\rm WD}$ and $\Omega_{\rm K} \equiv \sqrt{G M_1/{R_1}^3}$ denote the angular velocity of the WD rotation and the Keplerian angular velocity of the accreting gas.
Here, $\dot{M}$ is mass accretion rate onto the WD, and $R_1$ represents the WD radius and we assume that $\Omega_{\rm WD} \ll \Omega_{\rm K}$.

When the BL is optically thick, this energy loss is converted into the blackbody radiation from an equatorial belt around the WD surface 
with the vertical scale height ($H$).
The effective temperature ($T_{\rm eff}$) of the emitting ring is determined by the following equation:
\begin{equation}
 \dot{E}_{\rm loss} = 4 \pi R_1 H~\sigma {T_{\rm eff}}^4.
\label{Teff-thick-bl}
\end{equation}
Since hydrostatic equilibrium is satisfied in the vertical direction, the following relation holds.
\begin{equation}
  H = \frac{c_{\rm s}}{\Omega_{\rm K}}.
\label{scale-height}
\end{equation}
Here, $c_{\rm s} \equiv \sqrt{\frac{k T}{\mu m_{\rm H}}}$ denotes the sound speed, where $k$, $\mu$, and $m_{\rm H}$ are the Boltzmann constant, the mean molecular weight, and the proton mass.
We adopt a mean molecular weight of 0.62 under the assumption that hydrogen and helium are ionized in the gas having the solar abundances.
Assuming the gas at the BL is isothermal with $T \approx T_{\rm eff}$, we can combine equations (\ref{Teff-thick-bl}) and (\ref{scale-height}) to solve for the effective temperature. 

The observed BL luminosity, $L_{\rm BL}$, depends on the geometry of the system. 
We assume that the emission originates from a cylindrical strip of height $H$ at the inner edge of the disk ($R_{\rm in} \approx R_1$). 
Accounting for the projection effect and the fact that the lower half of the BL is occulted by the optically thick accretion disk, \textcolor{black}{the observed luminosity} is expressed as:
\begin{equation}
L_{\rm BL} = 2 R_1 H \sigma T_{\rm eff}^4 \sin i,
\label{lumi-bl} 
\end{equation}
where $i$ represents the inclination angle. 
We calculate the EUV flux in the wavelength range 70--760~\AA\ to compare with the EUVE observations.

\subsection{Calculation of X-ray Luminosity during quiescence}\label{ssec:34}

During quiescence, the mass accretion rate onto the WD remains low. 
Consequently, the inner accretion flow is no longer optically thick; instead, the gas evaporates into a high-temperature, optically thin, and geometrically thick coronal flow \citep{mey94siphonflow}. 
This coronal gas eventually accretes onto the WD surface, forming a BL.
As discussed in the previous subsection, the BL is optically thin during quiescence, and its radiation is dominated by bremsstrahlung (free--free) X-ray emission. Both the coronal flow and the BL serve as the primary X-ray emitters.

Two possible configurations for the inner accretion flow during quiescence may be considered. 
In the first scenario, the energy generated by viscous heating in the coronal flow is efficiently radiated locally as X-ray emission. 
The coronal gas flow accretes onto the WD by losing its angular momentum via viscous torques. 
In this case, an optically thin BL forms a belt-like region on the WD surface (see the middle panel of Figure~\ref{fig:inner-flow-picture}).
In the second scenario, the coronal flow is radiatively inefficient, resembling an advection-dominated accretion flow \citep[ADAF;][]{nar94ADAF}, and accretes onto the WD on dynamical timescales. 
In this case, most of the dissipated energy is advected onto the WD surface and radiated in the BL. 
The BL is expected to cover the entire surface of the WD as the accreting gas flow becomes more or less spherical (see the right panel of Figure~\ref{fig:inner-flow-picture}).

Detailed modeling of these scenarios is beyond the scope of this paper and will be addressed in future work. Instead, we estimate the maximum possible X-ray luminosity using the following expression:
\begin{equation}
L_{X} = (1-\lambda) GM_{1}\dot{M}_{\rm evap}
\left( \frac{1}{R_1} - \frac{1}{2 R_{\rm in}} \right).
\label{X-ray-flux}
\end{equation}
Here, we assume that both the gravitational potential energy released between $R_{\rm in}$ and $R_{\rm WD}$ and the rotational energy are fully thermalized and emitted as X-ray radiation. 
The radius $R_{\rm in}$ denotes the outer boundary of the coronal flow and is identical to the inner edge of the accretion disk. 
We note that the X-ray luminosity obtained from equation~(\ref{X-ray-flux}) represents an upper limit and can exceed the observed X-ray luminosity measured in a specific energy band.

\section{Results}\label{sec:4}

\subsection{Binary parameters and observational quantities of SS Cyg and U Gem}\label{ssec:41}

\begin{table*}[htb]
  \tbl{Summary of the binary parameters and observational characteristics of SS Cyg and U Gem.}{%
  \begin{tabular}{|c|ccc|}
      \hline
      Parameters & SS Cyg & U Gem & Refs.\footnotemark[$\S\S$]\\ 
      \hline
      $M_1$~[$M_{\odot}$] & 0.94 & 1.20 & 1, 2 \\
      $q$\footnotemark[$*$] & 0.628 & 0.35 & 1, 2 \\
      $P_{\rm orb}$~[d]\footnotemark[$\dag$] & 0.27512973 & 0.176906 & 3, 4 \\
      $R_1$~[cm] & 5 $\times$ 10$^{8}$ & 4 $\times$ 10$^{8}$ & 5 \\
      $a$~[cm]\footnotemark[$\ddag$] & 1.43 $\times$ 10$^{11}$ & 1.09 $\times$ 10$^{11}$ & 1, 2 \\
      $R_{\rm tidal}$~[$a$]\footnotemark[$\S$] & 0.356 & 0.408 & 6 \\
      $R_{\rm LS}$~[$a$]\footnotemark[$\P$] & 0.105 & 0.134 & 7 \\
      $T_1$~[K]\footnotemark[$\|$] & 50,000 & 35,000 & 8, 9, 10 \\
      $T_2$~[K]\footnotemark[$\sharp$] & 4,200 & 3,100 & 1, 2 \\
      $i$~[deg] & 45 & 69.7 & 1, 2 \\
      $d$~[pc]\footnotemark[$**$] & 114.25 & 93.4 & 11 \\
      Outburst interval~[d] & $\sim$40 & $\sim$120 & 12, 13 \\
      Outburst amplitude~[mag] & $\sim$3.5 & $\sim$4.5 & 12, 13 \\
      Outburst duration~[d] & $\sim$15 & $\sim$10 & 12, 13 \\
      UV delay~[d] & $\sim$1.5 & $\sim$1.25 & 14, 15, 16 \\
      $V$ magnitude~[mag]\footnotemark[$\dag\dag$] & 8.2--11.7 & 9.5--14 & 12, 13 \\
      $L_{\rm X}$~[ergs~s$^{-1}$]\footnotemark[$\ddag\ddag$] & $\sim$3 $\times$ 10$^{32}$ & $\sim$3 $\times$ 10$^{31}$ & 10, 17 \\
      \hline
    \end{tabular}}\label{tab:first}
\begin{tabnote}
\footnotemark[$*$] The mass ratio of a binary system defined as $q \equiv M_2 / M_1$, where $M_2$ is the mass of the secondary star.\\
\footnotemark[$\dag$] The orbital period of the binary system. \\ 
\footnotemark[$\ddag$] The binary separation. \\
\footnotemark[$\S$] The tidal truncation radius. \\
\footnotemark[$\P$] The Lubow-Shu radius. \\  
\footnotemark[$\|$] The WD effective temperature. \\ 
\footnotemark[$\sharp$] The effective temperature of the secondary star. \\ 
\footnotemark[$**$] The distance to the target from the earth. \\
\footnotemark[$\dag\dag$] The range of $V$-band magnitude from outbursts to the quiescent state. \\ 
\footnotemark[$\ddag\ddag$] The X-ray luminosity in the quiescent state in 0.4--10~keV for SS Cyg and in 1.0--7.0~keV for U Gem. \\ 
\footnotemark[$\S\S$] References: 1.~\citet{hil17sscyg}, 2.~\citet{ech07ugem}, 3.~\citet{hes84sscyg}, 4.~\citet{mar90ugem}, 5.~\citet{nau72WDmassradius}, 6.~\citet{pac77diskmodel}, 7.~\citet{lub75AD}, 8.~\citet{sio10sscygWD}, 9.~\citet{fro01ugem}, 10.~\citet{guv06ugemX}, 11.~\citet{bai18gaia}, 12.~\citet{can98sscyg}, 13.~$<$https://www.aavso.org/LCGv2/$>$, 14.~\citet{whe03sscyg}, 15.~\citet{lon96ugem}, 16.~\cite{mau02euve}, 
17.~\citet{ish09sscygSuzaku}.\\
\end{tabnote}
\label{binary-parameters}
\end{table*}

The binary parameters of SS~Cyg and U~Gem, along with the observational quantities relevant to our simulations, are summarized  in Table~\ref{binary-parameters}.
Our numerical code includes several free parameters, which are summarized in Table~\ref{model-parameters}.
We determine the values for $\alpha_{\rm hot}$, the ratio $\alpha_{\rm cold}/\alpha_{\rm hot}$, and the mass-transfer rate $\dot{M}_{\rm tr}$ by fitting the model to reproduce the observed outburst amplitudes, durations, and recurrence intervals.
In these simulations, $\alpha_{\rm cold}$ is assumed to be radially constant.
The parameter $dr_{\rm s}$, \textcolor{black}{the radial width of the outermost part of the disk in which the gas stream from the secondary star is transferred}, controls the number of short outbursts between two long outbursts.
We fix this parameter at $0.02a$, following \citet{kim23sscyg}, for SS~Cyg and at $0.03a$ for U~Gem.
We set \textcolor{black}{the coefficient of the tidal torque $c\omega$ at 0.4~s$^{-1}$} based on the prescription of \citet{ich94tidal}, \textcolor{black}{where $c$ and $\omega$ are the constant and the angular velocity of the binary motion, respectively}.

The degree of evaporation is determined by the parameters $\dot{M}_{\rm base}$ and $\lambda$ in equation~(\ref{mdot-evap}).
We adopt the standard values $\dot{M}_{\rm base} = 10^{15}$~g~s$^{-1}$ and $\lambda$ = 0.25, following \citet{mey94siphonflow}.
However, as discussed in subsection \ref{sec:51}, the standard evaporation rate is insufficient to account for the high X-ray luminosities observed during the quiescence of SS~Cyg. 
Consequently, we also examine a case with an enhanced evaporation rate using $\dot{M}_{\rm base} = 10^{15.5}$~g~s$^{-1}$. 

\begin{table*}[htb]
  \tbl{Summary of the model parameters in our simulations of SS Cyg and U Gem.}{%
  \begin{tabular}{|c|cc|}
      \hline
      Parameters & SS Cyg & U Gem \\ 
      \hline
      $\alpha_{\rm hot}$\footnotemark[$*$] & 0.15 & 0.30 \\
      $\alpha_{\rm cold}$/$\alpha_{\rm hot}$~[d]\footnotemark[$\dagger$] & 0.1 & 0.1 \\
      $\log_{10}\dot{M}_{\rm tr}$~[g~s$^{-1}$]\footnotemark[$\ddagger$] & 16.9 & 16.2 \\
      $dr_{\rm s}$~[$a$] & 0.02 & 0.03 \\
      $c\omega$~[$s^{-1}$] & 0.4 & 0.4 \\
      $\dot{M}_{\rm base}$~[g~s$^{-1}$]\footnotemark[$\S$] & 10$^{15}$ & 10$^{15}$ \\
      $\lambda$ & 0.25 & 0.25 \\
      \hline
    \end{tabular}}\label{tab:first}
\begin{tabnote}
\footnotemark[$*$] The viscosity parameter value in the hot state.\\
\footnotemark[$\dagger$] The ratio of the viscosity parameter value in the cold branch to that in the hot branch. \\ 
\footnotemark[$\ddagger$] The mass transfer rate from the secondary star to the accretion disk in units of g~s$^{-1}$. \\ 
\footnotemark[$\S$] A parameter characterizing the strength of evaporation in units of g~s$^{-1}$. \\ 
\end{tabnote}
\label{model-parameters}
\end{table*}

\subsection{Numerical simulations of SS Cyg}\label{ssec:42}

\begin{figure*}
 \begin{center}
 \begin{minipage}{0.49\hsize}
  \includegraphics[width=8cm]{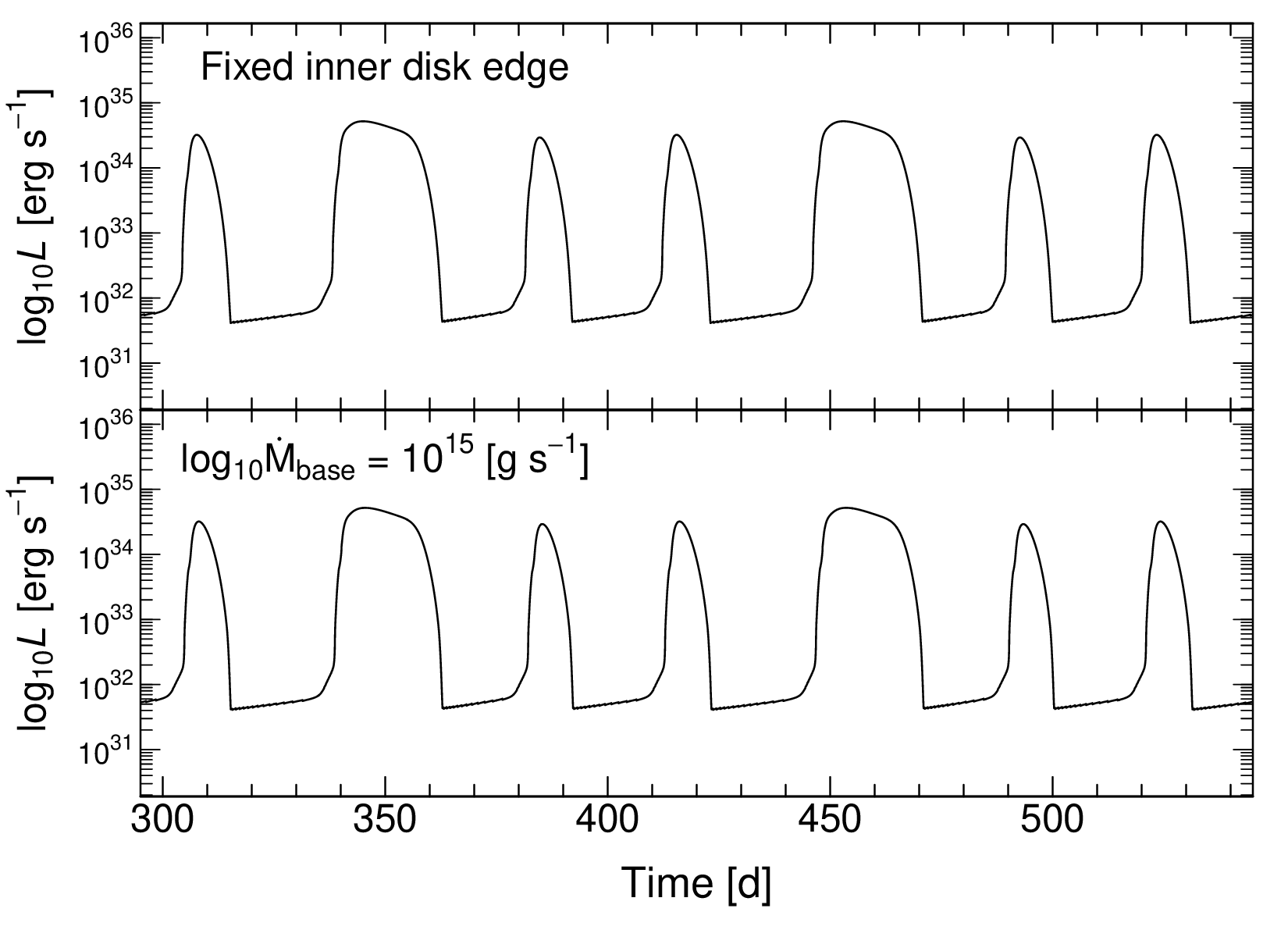} 
\end{minipage}
\begin{minipage}{0.49\hsize}
  \includegraphics[width=8cm]{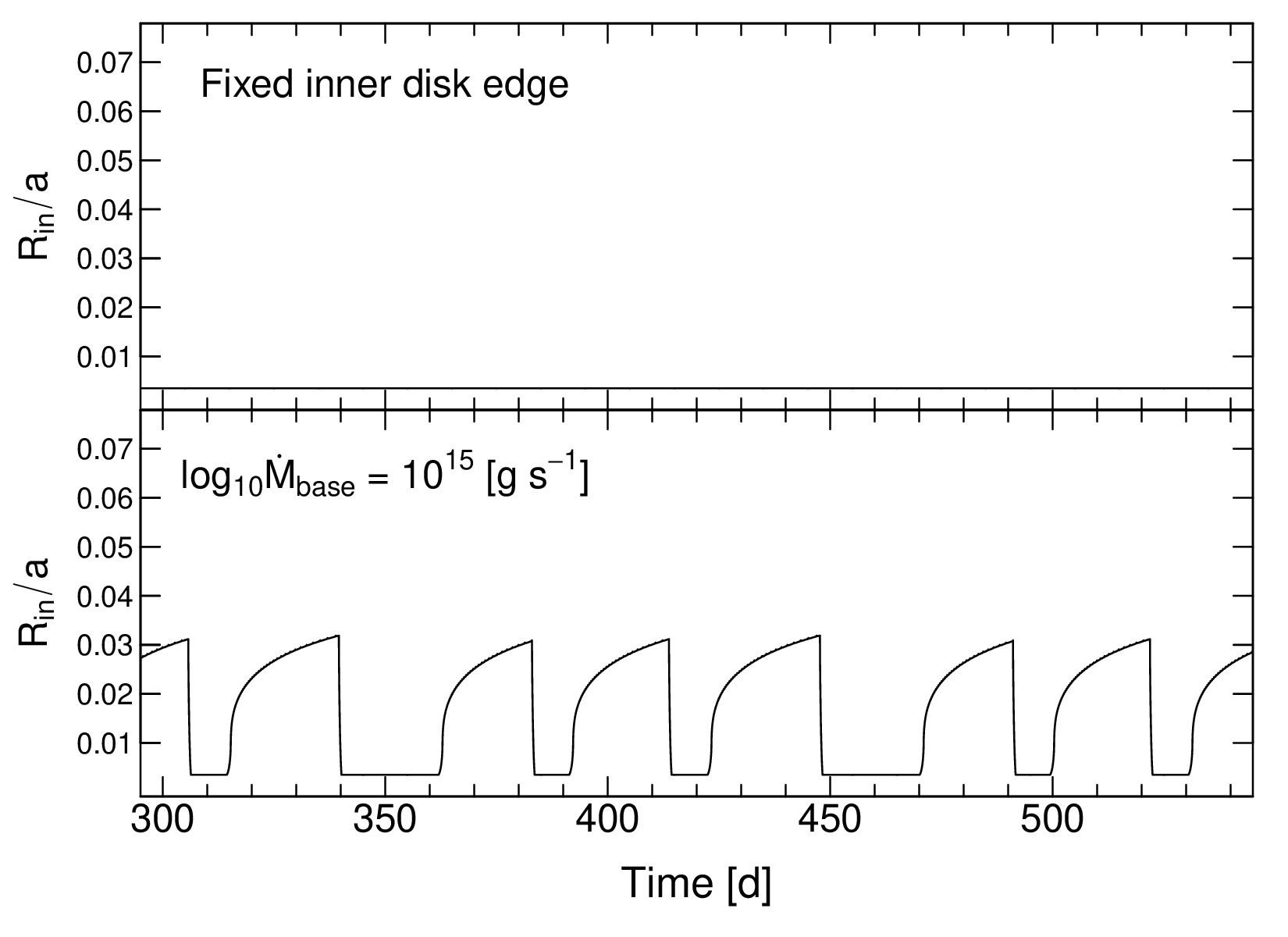}
\end{minipage}
\end{center}
\caption{
Time evolution of \textcolor{black}{the bolometric disk luminosity} (the left panels) and the inner disk radius (the right panels) for SS Cyg simulations.
The top row displays the model with a fixed inner disk edge, while the bottom row shows the model including evaporation with $\dot{M}_{\rm base} = 10^{15}$~g~s$^{-1}$.
{Alt text: Two columns of panels. X axis shows the time from 295 to 545 days. In the left figure, y axis shows \textcolor{black}{the bolometric disk luminosity} on a logarithmic scale from 10 to 30.5 to 10 to 36 ergs per second. In the right figure, y axis shows the inner disk edge in units of the binary separation from 0 to 0.075.}
 }
\label{fig:sscyg-3panels}
\end{figure*}

\begin{figure*}
 \begin{center}
 \begin{minipage}{0.49\hsize}
  \includegraphics[width=8cm]{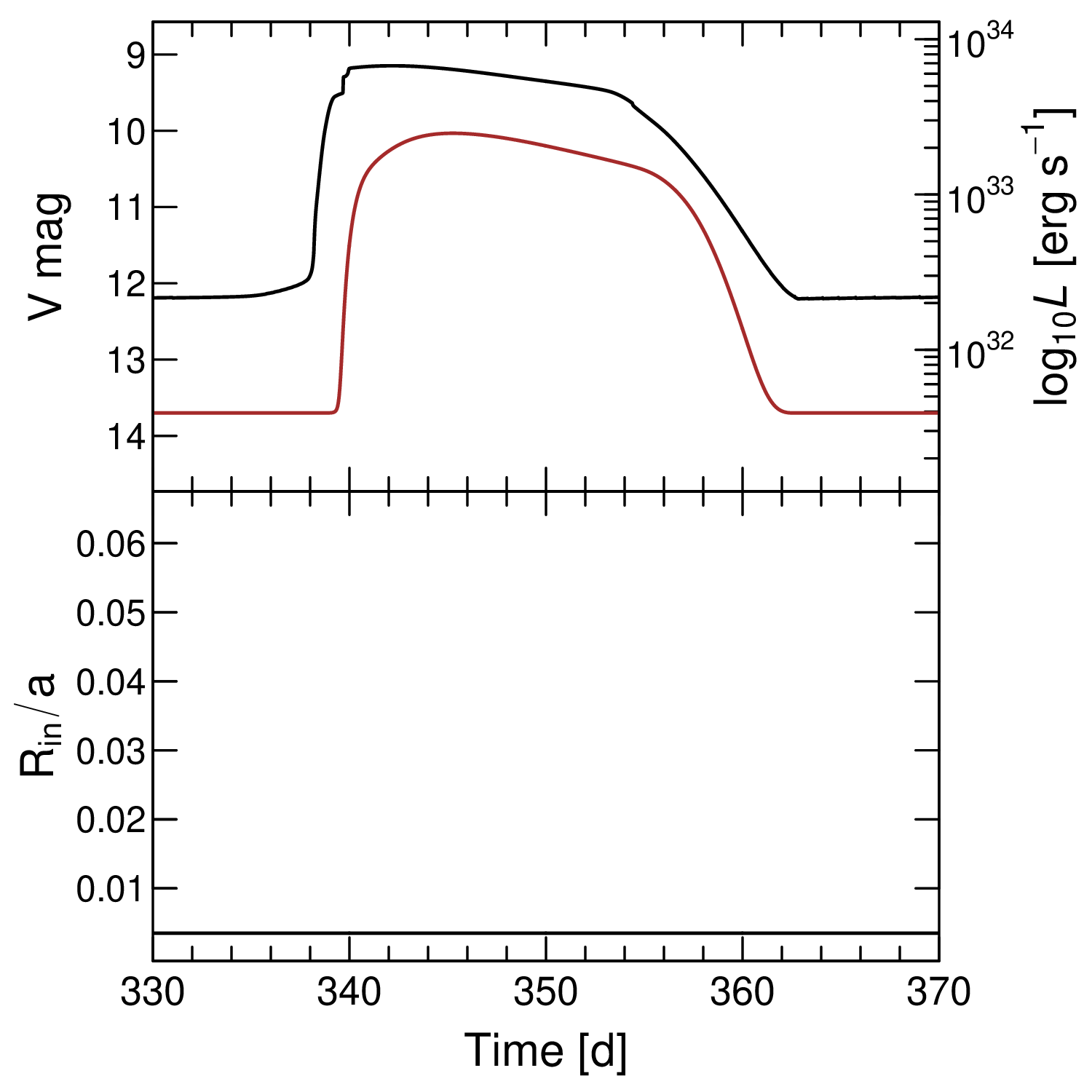} 
\end{minipage}
\begin{minipage}{0.49\hsize}
  \includegraphics[width=8cm]{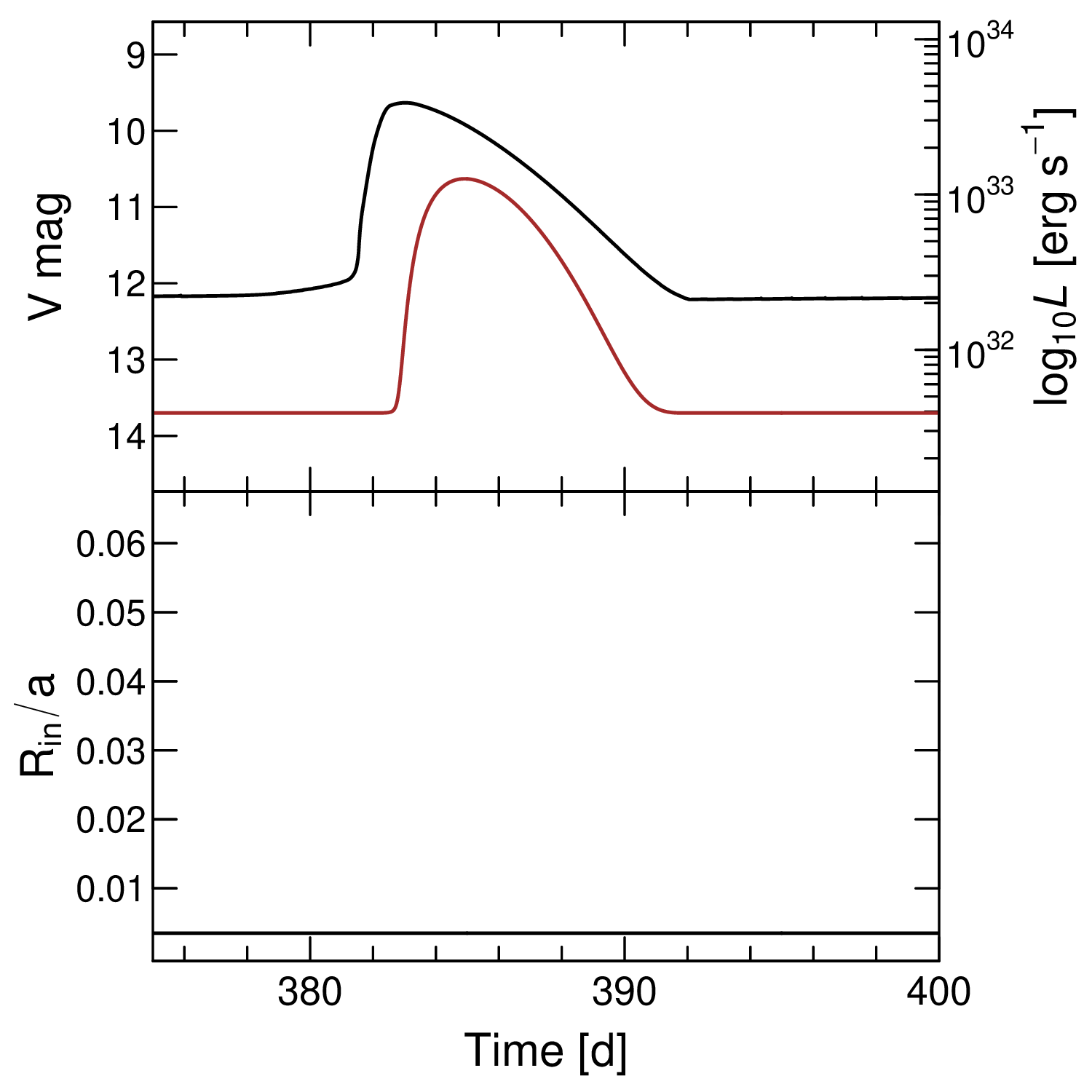}
\end{minipage}
\end{center}
\caption{\textcolor{black}{Enlarged views of a long outburst and a short outburst, which are produced by our simulations for SS Cyg without evaporation.
The upper panel shows the $V$-band magnitude (black) and the UV luminosity (brown).
The lower panel displays the inner disk radius.
{Alt text: Two columns of panels. In the left plot, x axis shows 330 to 370 days. In the right plot, x axis shows 375 to 400 days.}}
}
\label{fig:sscyg-noevap-long-short}
\end{figure*}

We first compare the light curves obtained from our numerical simulations with and without the effect of evaporation.
Figure~\ref{fig:sscyg-3panels} shows the light curves of \textcolor{black}{the bolometric disk luminosity} and the time evolution of the inner disk radius for the two cases.
Since SS~Cyg has a relatively high mass-transfer rate, inside-out outbursts do not occur, even when the inner disk radius is fixed at the WD surface.
Consequently, the light curves produced by the two simulations in Figure~\ref{fig:sscyg-3panels} appear nearly identical. 
However, significant discrepancies arise when compared to observational data. 
\textcolor{black}{Without evaporation, the calculated UV delay is $\sim$1~d, which is insufficient to account for the observed delay (see Table~\ref{model-parameters} and Figure \ref{fig:sscyg-noevap-long-short})}. 
Furthermore, the quiescent X-ray luminosity predicted with a fixed inner disk edge is more than two orders of magnitude lower than the observed values.

\begin{figure*}
 \begin{center}
  \includegraphics[width=12cm]{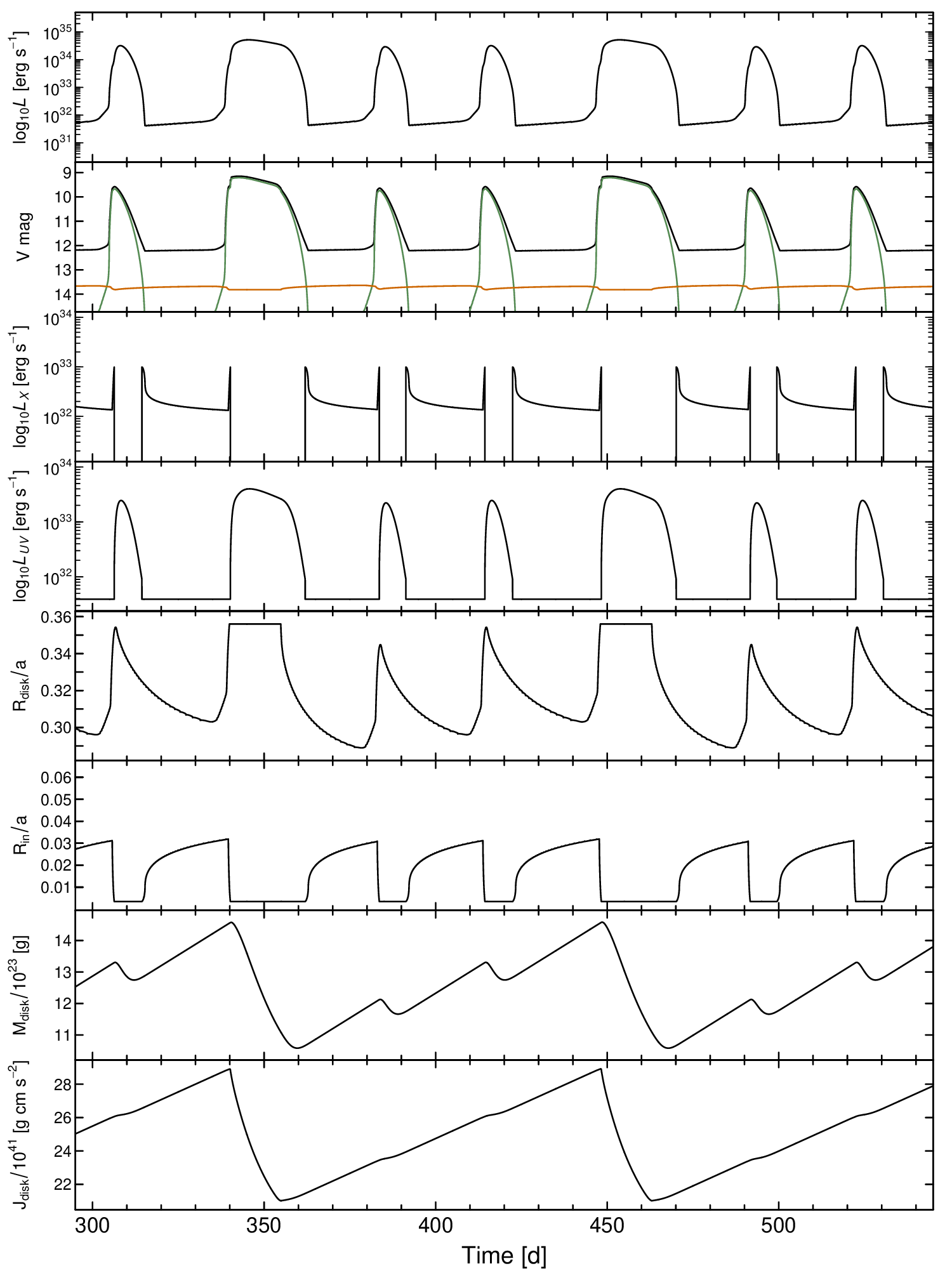} 
 \end{center}
\caption{Results of our simulation for SS Cyg with $\dot{M}_{\rm base} = 10^{15}$~g~s$^{-1}$.
From top to bottom: the light curve for \textcolor{black}{the bolometric disk luminosity} of the system, that for the $V$-band magnitude, the outer and inner disk radii in units of the binary separation, the total disk mass, and the
total disk angular momentum. 
The green and orange lines in the second panel represent the $V$-band magnitude of the disk and the bright spot, respectively. 
The black line in the same panel stands for a sum of the flux from these two components plus the WD and the secondary star.
The $V$-band magnitude of the secondary star and that of the WD are 12.8 and 14.8, respectively.
{Alt text: A single plot. X axis shows the time from 295 days to 545 days. 
In the second panel, y axis shows the V band magnitude from 14.5 to 8.8. 
In the third panel, y axis shows the X ray luminosity on a logarithmic scale from 10 to 31.2 to 10 to 34 ergs per second. 
In the fourth panel, y axis shows the UV luminosity on a logarithmic scale from 3 times 10 to 31 to 10 to 34 ergs per second. 
}
}
\label{fig:sscyg-best}
\end{figure*}

\begin{figure*}
 \begin{center}
 \begin{minipage}{0.49\hsize}
  \includegraphics[width=8cm]{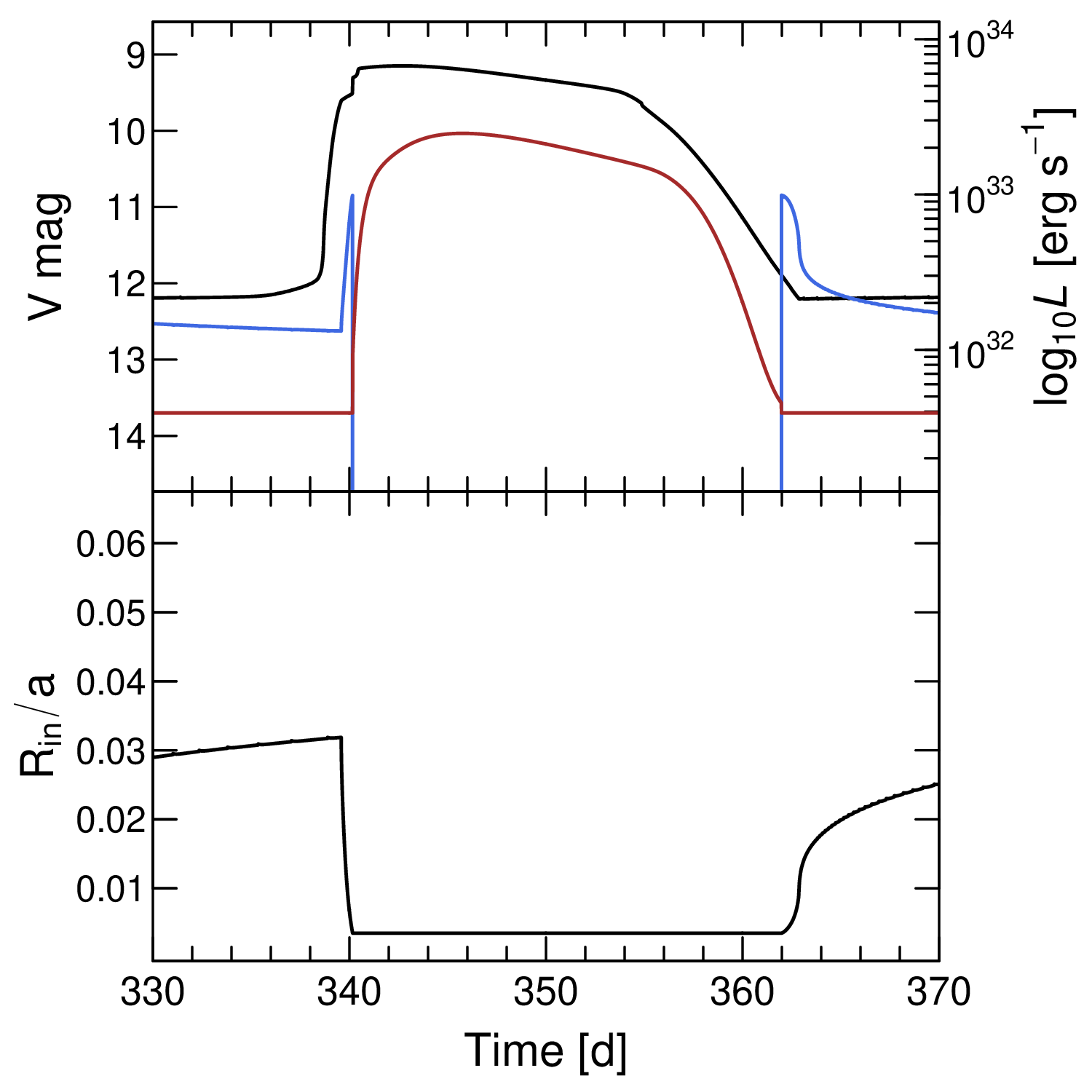} 
\end{minipage}
\begin{minipage}{0.49\hsize}
  \includegraphics[width=8cm]{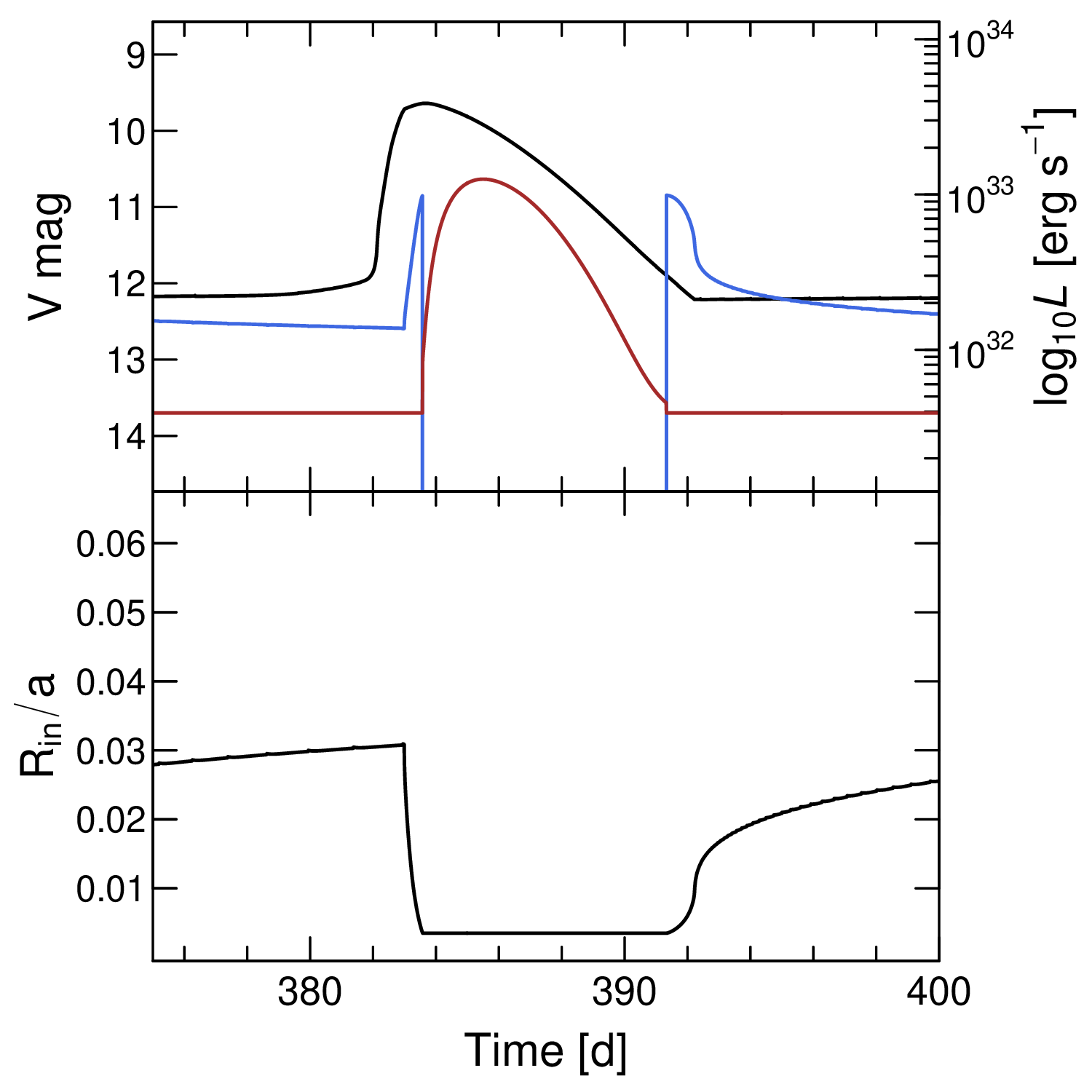}
\end{minipage}
\end{center}
\caption{Enlarged views of a long outburst and a short outburst, which are produced by our simulations for SS Cyg \textcolor{black}{with $\dot{M}_{\rm base} = 10^{15}$~g~s$^{-1}$}.
The upper panel shows the $V$-band magnitude (black), the X-ray luminosity (blue), the UV luminosity (brown).
The lower panel displays the inner disk radius.
{Alt text: Two columns of panels. In the left plot, x axis shows 330 to 370 days. In the right plot, x axis shows 375 to 400 days.}
}
\label{fig:sscyg-best-long-short}
\end{figure*}

Figure~\ref{fig:sscyg-best} displays the overall simulation results, while Figure~\ref{fig:sscyg-best-long-short} provides an enlarged view of the time evolution of the optical, UV, and X-ray luminosities, alongside the movement of the inner disk edge during both a long and a short outburst.
Upon the trigger of an outburst, the inner disk edge rapidly fills the evaporated hole, reaching the WD surface within approximately two days. 
This inward migration results in a delay of $\sim$2~d between the rise of the UV flux and the onset of the optical outburst, a value that is consistent with observations (see Table~\ref{binary-parameters}).
On the other hand, the predicted maximum X-ray luminosity during quiescence remains approximately a few times lower than the observed value. 
This problem will be discussed in subsection \ref{sec:51}.
Throughout the duration of the outburst, the inner disk edge stays at the WD surface. 
As the outburst subsides, the edge recedes from the WD surface, and the \textcolor{black}{coronal cavity} gradually expands throughout the quiescent phase. 
The rate of this expansion decreases over time. 
During quiescence, the median values for the inner disk radius and the mass accretion rate onto the WD are $R_{\rm in} = 3.9 \times 10^{9}$ cm and $\dot{M} = 8.8 \times 10^{14}$ g s$^{-1}$, respectively.

\subsection{Numerical simulations of U Gem}\label{ssec:43}

\begin{figure*}
 \begin{center}
 \begin{minipage}{0.49\hsize}
  \includegraphics[width=8cm]{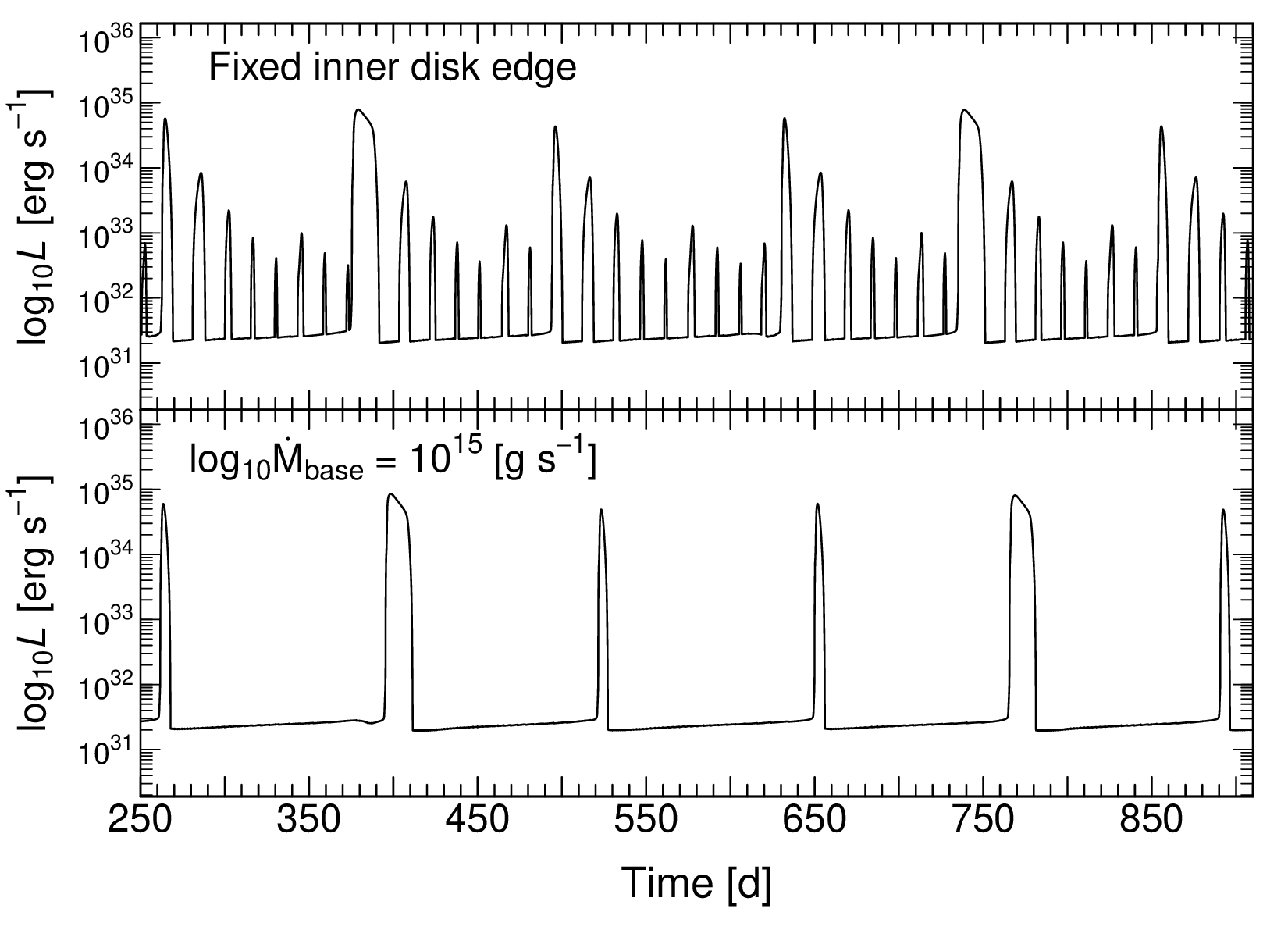} 
\end{minipage}
\begin{minipage}{0.49\hsize}
  \includegraphics[width=8cm]{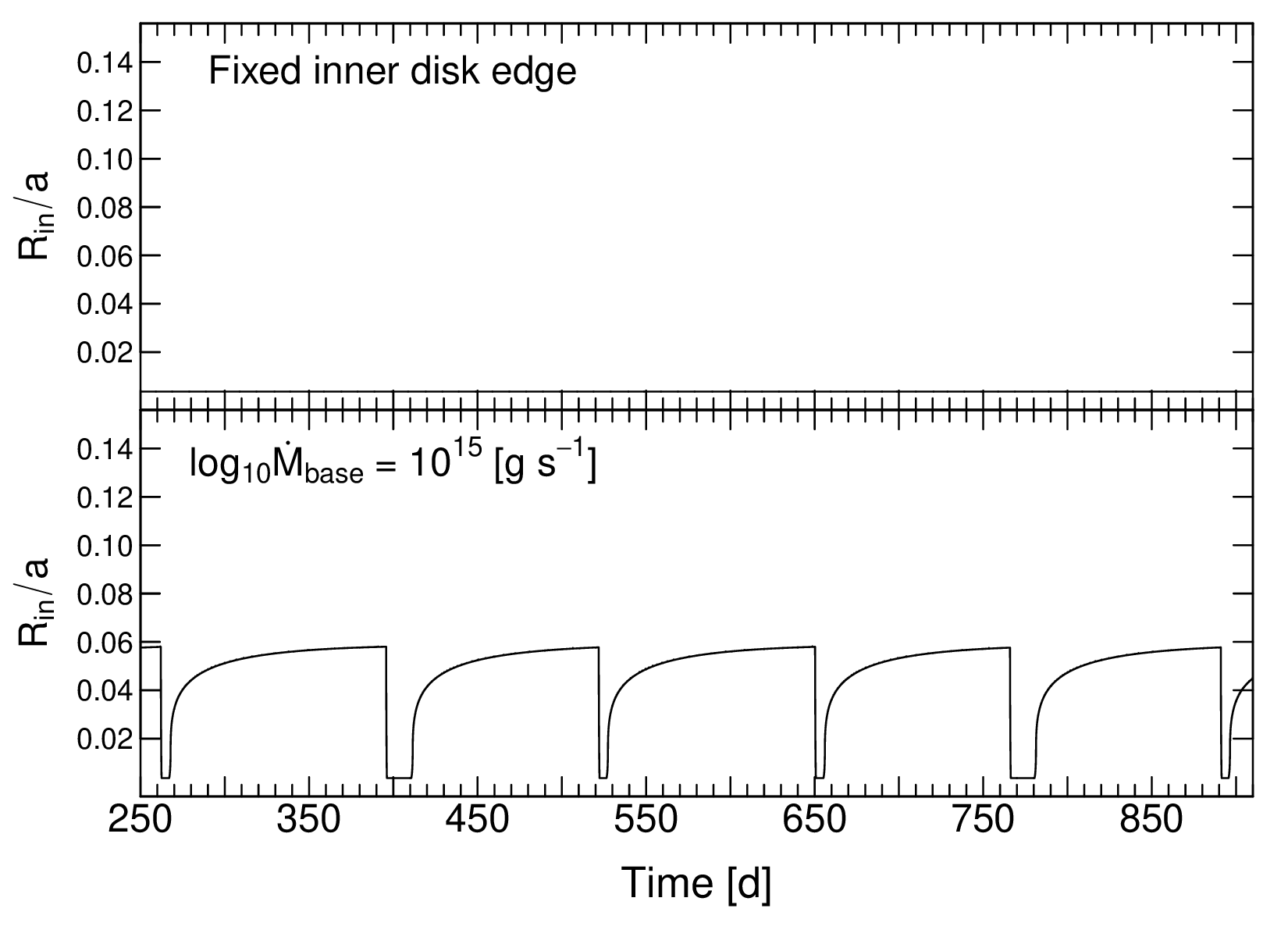}
\end{minipage}
\end{center}
\caption{Time evolution of \textcolor{black}{the bolometric disk luminosity} (the left panels) and the inner disk radius (the right panels) for U Gem simulations.
The top row displays the model with a fixed inner disk edge, while the bottom row shows the model including evaporation with $\dot{M}_{\rm base} = 10^{15}$~g~s$^{-1}$.
{Alt text: Two columns of panels. X axis shows the time from 295 to 910 days. In the left figure, y axis shows \textcolor{black}{the bolometric disk luminosity} on a logarithmic scale from 10 to 30.5 to 10 to 36 ergs per second. In the right figure, y axis shows the inner disk edge in units of the binary separation from 0 to 0.15.}
}
\label{fig:ugem-3panels}
\end{figure*}

As with SS~Cyg, we compare \textcolor{black}{the bolometric disk luminosity} light curves obtained from simulations with and without evaporation (see Figure~\ref{fig:ugem-3panels}). 
As shown in the top left panel of Figure~\ref{fig:ugem-3panels}, low-amplitude ``inside-out'' outbursts occur frequently between large-amplitude outbursts when the inner disk edge is fixed at the WD surface. 
This morphological difference arises because the mass transfer rate in U~Gem is significantly lower than that in SS~Cyg. 
In low-$\dot{M}$ systems, the surface density $\Sigma$ can reach the critical threshold $\Sigma_{\rm max}$, \textcolor{black}{above which the disk gas becomes thermally unstable}, in the inner disk before the outer disk, triggering an inside-out transition.
Furthermore, the maximum X-ray luminosity during quiescence predicted by the fixed-edge model is significantly lower than observational data. 
These discrepancies, both in outburst behavior and in quiescent emission, suggest that disk evaporation is a necessary component for reproducing the observed behavior of U~Gem.

\begin{figure*}
 \begin{center}
  \includegraphics[width=12cm]{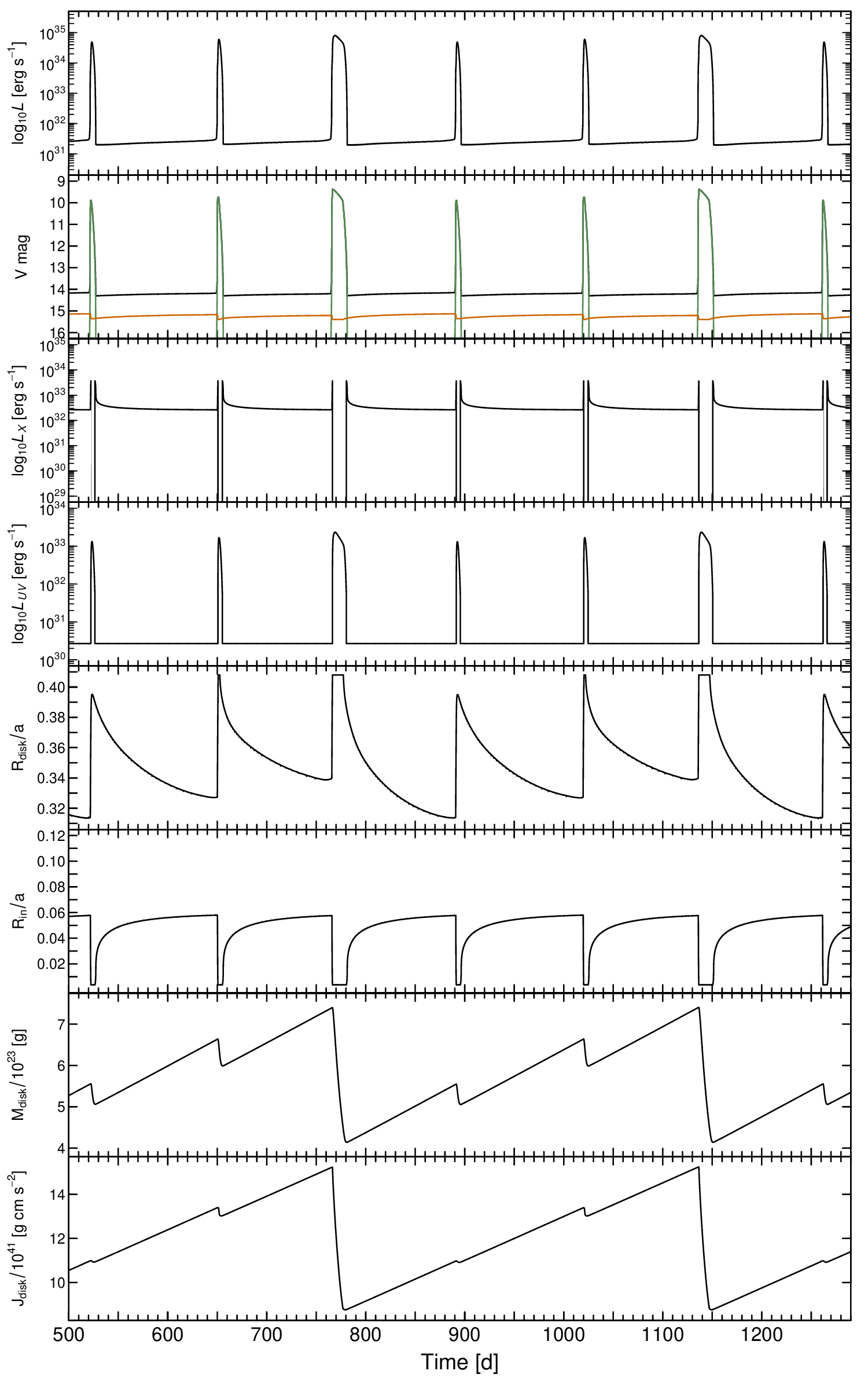} 
 \end{center}
\caption{Results of our simulation for U Gem with $\dot{M}_{\rm base} = 10^{15}$~g~s$^{-1}$.
From top to bottom: the light curve for \textcolor{black}{the bolometric disk luminosity}, that for the $V$-band magnitude, the outer and inner disk radii in units of the binary separation, the total disk mass, and the
total disk angular momentum. 
The green and orange lines in the second panel represent the $V$-band magnitude of the disk and the bright spot, respectively. 
The black line in the same panel stands for a sum of the flux from these two components plus the WD and the secondary star.
The $V$-band magnitude of the secondary star and that of the WD are 15.7 and 15.6, respectively.
{Alt text: A single plot. X axis shows the time from 500 days to 1290 days. 
In the second panel, y axis shows the V band magnitude from 16 to 9. 
In the third panel, y axis shows the X ray luminosity on a logarithmic scale from 10 to 29 to 10 to 35 ergs per second. 
In the fourth panel, y axis shows the UV luminosity on a logarithmic scale from 3 times 10 to 30 to 10 to 34 ergs per second. 
The light curve includes 7 outbursts.
The number of short outbursts between two long outbursts is 2.
}
}
\label{fig:ugem-best}
\end{figure*}

\begin{figure}
 \begin{center}
 \begin{minipage}{0.49\hsize}
  \includegraphics[width=8cm]{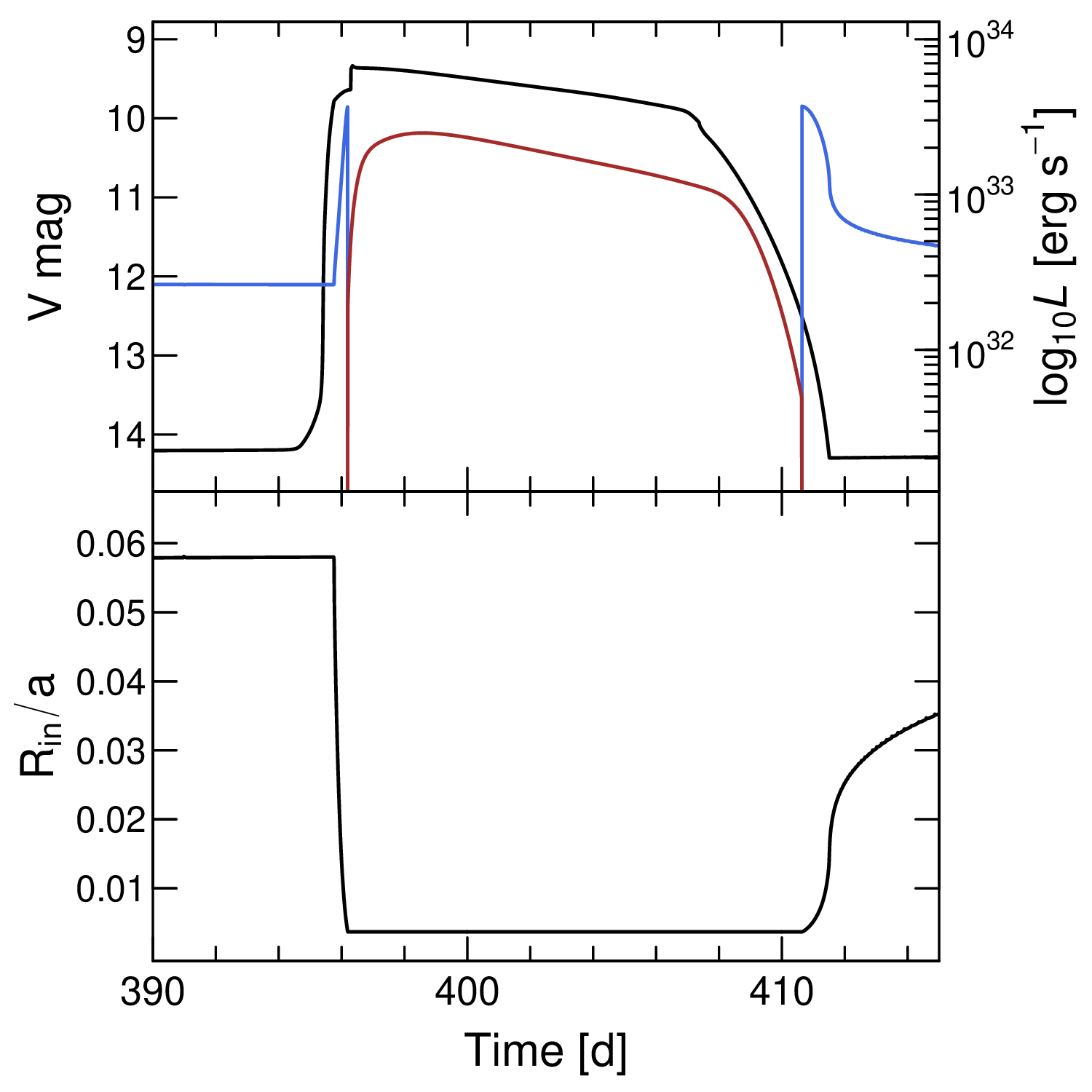} 
\end{minipage}
\begin{minipage}{0.49\hsize}
  \includegraphics[width=8cm]{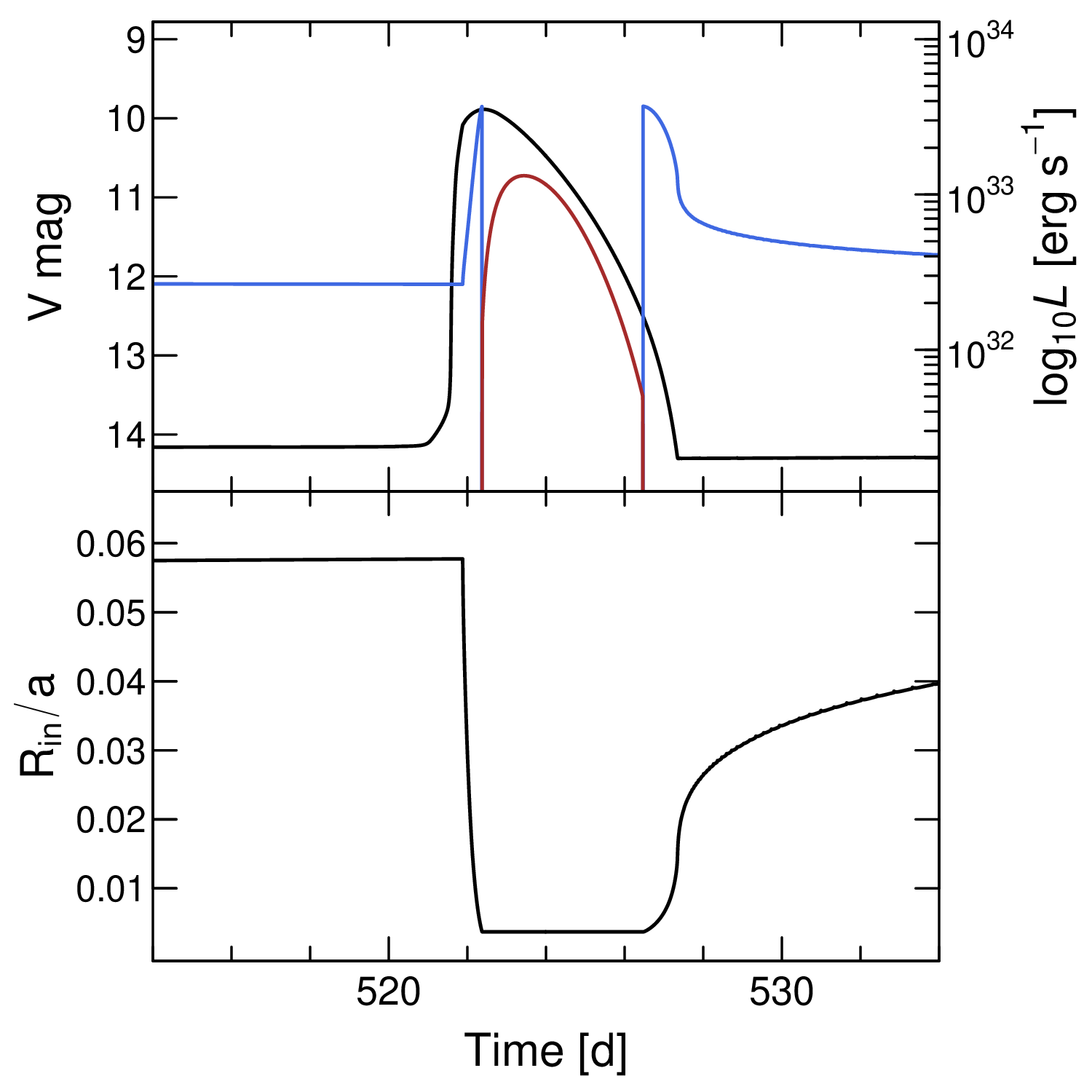}
\end{minipage}
\end{center}
\caption{Enlarged views of a long outburst and a short outburst, which are produced by our simulations for U Gem \textcolor{black}{with $\dot{M}_{\rm base} = 10^{15}$~g~s$^{-1}$}.
The upper panel shows the $V$-band magnitude (black), the X-ray luminosity (blue), the UV luminosity (brown).
The lower panel displays the inner disk radius.
{Alt text: Two columns of panels. In the left plot, x axis shows 390 to 415 days. In the right plot, x axis shows 514 to 534 days.}
}
\label{fig:ugem-best-long-short}
\end{figure}

Figure~\ref{fig:ugem-best} displays the overall simulation results, while Figure~\ref{fig:ugem-best-long-short} provides a detailed view of the time evolution of the optical, UV, and X-ray luminosities, alongside the movement of the inner disk edge during both long and short outbursts.
The fundamental characteristics are consistent with those identified in our simulations of SS~Cyg. 
Specifically, the delay between the rise of the UV flux and the onset of the optical outburst is approximately $1$~d. 
This delay is shorter than that in the case of SS~Cyg simulation, yet it remains in good agreement with observational data (see Table~\ref{binary-parameters}).
Furthermore, the X-ray luminosity derived from our simulations exceeds the observed X-ray luminosity of U~Gem. This is expected, as the luminosity calculated in this study represents an upper limit. 
A more realistic estimate is anticipated to be lower once detailed modeling of the optically thin BL and the coronal flow is performed, incorporating geometric effects and radiative efficiencies. 
During the quiescent phase, the median values for the inner disk radius and the mass accretion rate are $R_{\rm in} = 5.9 \times 10^{9}$~cm and $\dot{M} = 7.3 \times 10^{14}$~g~s$^{-1}$, respectively.

\section{Discussion}\label{sec:5}

\subsection{Requirements to account for the quiescent X-ray luminosity in SS~Cyg}\label{sec:51}

Our simulations indicate that disk truncation is essential to account for both the high quiescent X-ray luminosity and the $\gtrsim$1~d UV delay observed at outburst onset. 
However, the observed quiescent X-ray luminosity of SS~Cyg is exceptionally high; even our simulations based on the standard evaporation model \citep{mey94siphonflow} fail to reproduce these levels, as the resulting mass accretion rate remains insufficient. Conversely, this evaporation-based framework successfully matches other observables and the behavior of U~Gem.
To reconcile the high quiescent X-ray luminosity in SS~Cyg, we investigate two possible scenarios: (1) more efficient evaporation from the inner disk, or (2) a direct mass supply to the coronal cavity via mass overflow of the transferred gas beyond the outer disk edge.

\begin{figure*} 
\begin{center}
\includegraphics[width=12cm]{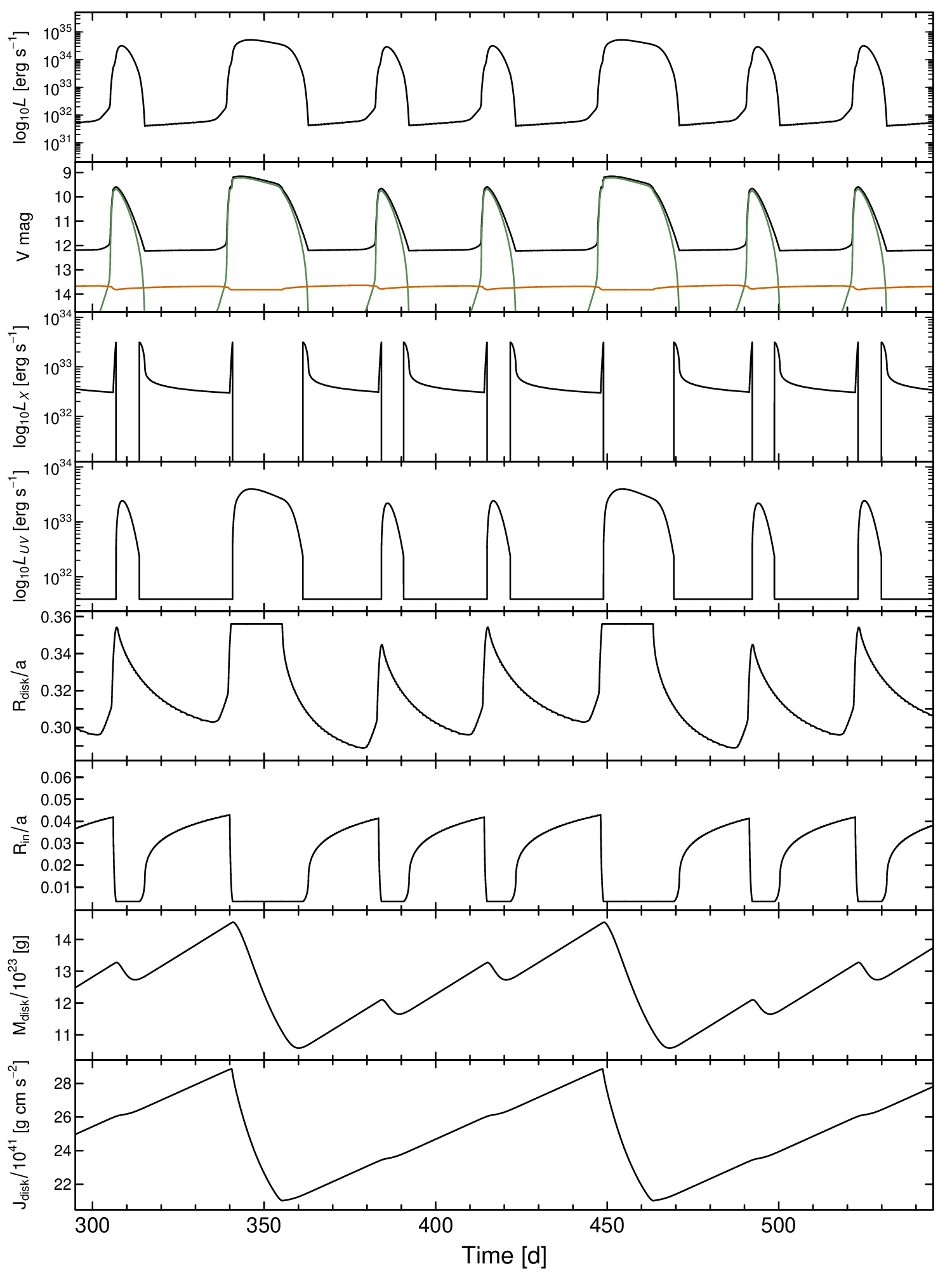} 
\end{center}
\caption{Same as Figure~\ref{fig:sscyg-best}, but showing the results of our simulation for SS~Cyg with an enhanced evaporation rate of $\dot{M}_{\rm base} = 10^{15.5}$~g~s$^{-1}$. From top to bottom: the light curve for \textcolor{black}{the bolometric disk luminosity}, that for the $V$-band magnitude, the outer and inner disk radii in units of the binary separation, the total disk mass, and the total disk angular momentum. The green and orange lines in the second panel represent the $V$-band magnitude of the disk and the bright spot, respectively. The black line in the same panel stands for a sum of the flux from these two components plus the WD and the secondary star. The $V$-band magnitude of the secondary star and that of the WD are 12.8 and 14.8, respectively.
{Alt text: A single plot. X axis shows the time from 295 days to 545 days. }
}
\label{fig:sscyg-mdb155}
\end{figure*} 

In the first scenario, a significantly enhanced evaporation rate is required. 
Since the X-ray luminosity in our simulation using the standard evaporation rate is several times lower than that observed in SS~Cyg, we examine a case with $\dot{M}_{\rm base} = 10^{15.5}$~g~s$^{-1}$. 
Figure~\ref{fig:sscyg-mdb155} illustrates the results for this case, which yields an X-ray luminosity slightly exceeding the observed value. 
Under these conditions, the median mass accretion rate and inner disk radius during quiescence are $\dot{M} = 2.4 \times 10^{15}$~g~s$^{-1}$ and $R_{\rm in} = 4.5 \times 10^{9}$~cm, respectively. 
\textcolor{black}{Whether the siphon-flow model by \citet{mey94siphonflow}, or any other evaporation model, can physically justify such a high evaporation rate for SS Cyg remains to be determined and is beyond the scope of this study}.

Regarding the second scenario, we previously suggested that the high X-ray luminosity in SS~Cyg could be explained if approximately 5\%  of the transferred gas overflows the outer disk and is supplied directly to the coronal flow \citep{kim23sscyg}. 
However, that estimate was derived from our numerical calculations based on a model with a fixed inner disk radius. 
When full evaporation is taken into account, we find that only $\sim$2\% of the transferred gas into the coronal cavity is sufficient to account for the observed X-ray luminosity in SS~Cyg. 
The outburst recurrence interval remains largely unaffected even when $\sim$2--5\% of the transferred gas overflows the disk.

\subsection{Interpretation of the time-varying X-ray luminosity}\label{sec:52}

As shown in Figures~\ref{fig:sscyg-best-long-short} and \ref{fig:ugem-best-long-short}, the X-ray luminosity exhibits a sharp spike followed by an abrupt drop to zero immediately preceding the rise in EUV luminosity. 
At first glance, this result appears consistent with the X-ray observations reported by \citet{whe03sscyg} (see their Figure 2). However, there is a significant caveat to this interpretation. 
\textcolor{black}{
The mass accretion rate increases rapidly once the heating front reaches the inner disk edge during an outburst. 
Our model assumes that the evaporation rate ($\dot{M}_{\rm evap}(R_{\rm in})$) instantaneously tracks the mass accretion rate at the inner disk edge ($\dot{M}_{\rm in}(R_{\rm in})$), as defined in equation (\ref{evaporation}). 
Physically, this assumption likely breaks down during such rapid transitions. 
We adopt the steady-state solution of \citet{mey94siphonflow} in our simulations, however, they noted that it typically takes several days, for the evaporation process to reach completion (see their equation (21)). 
Therefore, the evaporation rate cannot respond to variations on such short timescales.}
A similar phenomenon occurs at the end of the outburst when the inner disk edge recedes from the WD surface. 
\textcolor{black}{Consequently, although these numerical spikes should not be interpreted quantitatively, they nevertheless suggest rapid X-ray luminosity variations driven by changes in the mass accretion rate immediately before the EUV rise and after its decline.}

In contrast, the slower timescales characteristic of quiescence allow for a more reliable discussion of X-ray evolution. 
During these periods, the luminosity gradually decreases. 
As noted by \citet{mey94siphonflow}, VW Hyi exhibits a gradual decline in X-ray flux \citep{vanderwoe87vwhyi}, a behavior qualitatively reproduced by our simulations. 
Similarly, \citet{mcg04sscygXray} demonstrated that SS Cyg enters a slow X-ray decline following an outburst. 
In our results, the quiescent X-ray luminosity follows a power law of $\dot{M}_{\rm evap}(R_{\rm in}) \propto (t - t_0)^{-1/4}$ for SS Cyg and $\dot{M}_{\rm evap}(R_{\rm in}) \propto (t - t_0)^{-1/8}$ for U Gem.
Here, $t_0$ denotes the time at the end of an outburst. 
\textcolor{black}{These relations are derived by the fitting of the simulated X-ray light curve with an exponential function.} 
Notably, these decline rates are close to, but less steep than, the earlier simulation results of \citet{mey94siphonflow}.

The limitations regarding rapid temporal variations in the evaporation rate also extend to the evolution of the inner disk edge, $R_{\rm in}$, and the resulting UV delay. 
Because the physical evaporation process may not respond on these short timescales, the inward migration of the inner disk edge could be even more rapid than our models predict. 
\textcolor{black}{On the other hand, the outward migration could become more slowly.}
If the evaporation rate fails to increase instantaneously at the onset of an outburst, the UV delay might be shortened; however, this remains contingent on the timescale over which the accretion rate onto the BL increases. 
Ultimately, both the UV delay and the rapid motion of the inner disk edge should be interpreted with caution. 
Nevertheless, the inclusion of evaporation robustly leads to a longer UV delay compared to models that neglect the process entirely.

\section{Summary}\label{sec:6}

In this study, we performed numerical simulations of time-dependent accretion disks in SS~Cyg and U~Gem using a one-dimensional code that incorporates variable inner and outer disk edges. 
By accounting for the BL and calculating the X-ray luminosity, we modeled the multi-wavelength emission characteristics of these systems. 
The primary objective of this work was to investigate the behavior of the thermal-viscous instability in the presence of inner disk evaporation during quiescence and its impact on the resulting optical, UV, and X-ray light curves.
Our major findings are summarized as follows:
\begin{itemize}
\item To reproduce the observed emission characteristics, the inner disk edge must recede to several $\times 10^{9}$~cm during the quiescent phase (see subsections~\ref{ssec:42} and \ref{ssec:43}).
\item The presence of an inner cavity during quiescence suppresses the triggering of inside-out outbursts. This eliminates the need for a radial dependence of the $\alpha_{\rm cold}$ parameter, as the instability is physically prevented from initiating in the innermost disk region (see Figure~\ref{fig:ugem-3panels}).
\item The time delay between the rise in UV luminosity and the onset of the optical outburst is limited to approximately 0.5~d in the absence of evaporation. 
This delay increases significantly as a larger fraction of the inner disk evaporates into the coronal flow \textcolor{black}{(see subsections \ref{ssec:42} and \ref{ssec:43})}.
\item For U~Gem, disk evaporation supplies sufficient mass to the X-ray-emitting region to account for the observed quiescent X-ray luminosity (see Figure \ref{fig:ugem-best}).
\item In contrast, explaining the exceptionally high quiescent X-ray luminosity in SS~Cyg requires either significantly higher evaporation efficiency or a supplementary mass supply from gas overflowing the outer disk edge (see subsection~\ref{sec:51}).
\item The simulations reproduce a gradual decline in X-ray flux during quiescence, which is consistent with observational trends (see subsection~\ref{sec:52}).
\end{itemize}

The evaporation scenario provides a plausible physical mechanism for disk truncation in non-magnetic cataclysmic variables. 
While our simulations offer an updated framework for the disk instability model, further investigation is required to determine if the high accretion rates inferred for SS~Cyg can be fully reconciled within the evaporation framework. 
Future research will focus on more detailed numerical modeling of the coronal flow and the physics of the optically thin BL.

\begin{ack}
This work was financially supported by Japan Society for the Promotion of Science Grants-in-Aid for Scientific Research (KAKENHI) Grant Numbers JP21K13970 (MK) and JP24H01807 (MK).
We thank the anonymous referee for his or her insightful comments.
\end{ack}


\newcommand{\noop}[1]{}

\end{document}